\documentclass[12pt]{article}

\usepackage[margin=1in,nohead,dvips,a4paper]{geometry}
\usepackage{times,pstricks,pst-node}
\usepackage{amssymb}
\usepackage[small,center]{caption}

\newtheorem{theorem}{Theorem}
\newtheorem{lemma}[theorem]{Lemma}
\newtheorem{corollary}[theorem]{Corollary}

\makeatletter
\DeclareRobustCommand{\qed}{%
  \ifmmode 
  \else \leavevmode\unskip\penalty9999 \hbox{}\nobreak\hfill
  \fi
  \quad\hbox{\qedsymbol}}
\newcommand{\openbox}{\leavevmode
  \hbox to.77778em{%
  \hfil\vrule
  \vbox to.675em{\hrule width.6em\vfil\hrule}%
  \vrule\hfil}}
\newcommand{\qedsymbol}{\openbox}
\newenvironment{proof}[1][\proofname]{\par
  \normalfont
  \topsep6\p@\@plus6\p@ \trivlist
  \item[\hskip\labelsep\itshape
    #1.]\ignorespaces
}{%
  \qed\endtrivlist
}
\newcommand{\proofname}{Proof}
\makeatother

\def\bm #1{\mbox{\boldmath $#1$}}

\def\rmi {\mathrm i}
\def\Pf {\mathop{\mathrm{Pf}}}

\psset{linewidth=.5pt,dash=4pt 4pt,unit=.25in,arrowsize=2pt 3}

\def\arrowLine(#1,#2)(#3,#4){%
  \pcline(#1,#2)(#3,#4)%
  \lput{:U}{
    \pspicture(0,0)(0,0)
      \psline[arrows=->](2.3pt,0)(2.4pt,0)
    \endpspicture
  }
}

\def\vertexA{%
  \pspicture(0,0)(2,2)
    \arrowLine(0,1)(1,1)
    \arrowLine(1,1)(1,2)
    \arrowLine(2,1)(1,1)
    \arrowLine(1,1)(1,0)
  \endpspicture
}

\def\vertexB{%
  \pspicture(0,0)(2,2)
    \arrowLine(1,1)(0,1)
    \arrowLine(1,2)(1,1)
    \arrowLine(1,1)(2,1)
    \arrowLine(1,0)(1,1)
  \endpspicture
}

\def\vertexC{%
  \pspicture(0,0)(2,2)
    \arrowLine(0,1)(1,1)
    \arrowLine(1,1)(1,2)
    \arrowLine(1,1)(2,1)
    \arrowLine(1,0)(1,1)
  \endpspicture
}

\def\vertexD{%
  \pspicture(0,0)(2,2)
    \arrowLine(1,1)(0,1)
    \arrowLine(1,2)(1,1)
    \arrowLine(2,1)(1,1)
    \arrowLine(1,1)(1,0)
  \endpspicture
}

\def\vertexE{%
  \pspicture(0,0)(2,2)
    \arrowLine(1,1)(0,1)
    \arrowLine(1,1)(1,2)
    \arrowLine(2,1)(1,1)
    \arrowLine(1,0)(1,1)
  \endpspicture
}

\def\vertexF{%
  \pspicture(0,0)(2,2)
    \arrowLine(0,1)(1,1)
    \arrowLine(1,2)(1,1)
    \arrowLine(1,1)(2,1)
    \arrowLine(1,1)(1,0)
  \endpspicture
}

\def\uvertexLeft{%
  \pspicture[.4](0,-.1)(1.75,1.1)
    \psline{->}(0,0)(.5,0)
    \psline{-<}(.4,1)(.5,1)
    \psline(0,1)(.5,1)
    \psarc(.5,.5){.5}{270}{90}
    \qdisk(1,.5){.1}
    \rput(1.5,0.5){$x$}
  \endpspicture
}

\def\uvertexRight{%
  \pspicture[.4](0,-.1)(1.75,1.1)
    \psline{-<}(0,0)(.5,0)
    \psline(0,0)(.5,0)
    \psline{->}(0,1)(.5,1)
    \psarc(.5,.5){.5}{270}{90}
    \qdisk(1,.5){.1}
    \rput(1.5,0.5){$x$}
  \endpspicture
}

\begin{document}

\title{On refined enumerations\\
of some symmetry classes of ASMs}
\author{A.~V.~Razumov, Yu.~G.~Stroganov\\
\small \it Institute for High Energy Physics\\[-.5em]
\small \it 142280 Protvino, Moscow region, Russia}
\date{}

\maketitle

\begin{abstract}
Using determinant representations for partition functions of the
corresponding square ice models and the method proposed recently by one of
the authors, we investigate refined enumerations of vertically symmetric
alternating-sign matrices, off-diagonally symmetric alternating-sign
matrices and alternating-sign matrices with U-turn boundary.
For all these cases the explicit formulas for refined enumerations are
found. It particular, Kutin--Yuen conjecture is proved.
\end{abstract}

\section{Introduction} \label{s:1}

An alternating-sign matrix (ASM) is a matrix with entries 1, 0, $-1$, such
that 1's and $-1$'s alternate in each column and each row, and such that
the first and last non-zero entry in each row and column is 1. During last
decade many enumeration and equinumeration results related to the ASMs were
conjectured and proved. Nevertheless, there is a lot of problems to be
solved. Besides of their importance for pure combinatorics enumeration
results on ASMs will undoubtedly find numerous applications to problems of
mathematical physics, see in this respect, for example, papers
\cite{RSt01a, BGN01, RSt01b, RSt01c, PRG01, RSt01d, GBNM01, BGN02, GNPR03,
Zub03, FZZ03}.

In the present paper, using the method prposed by one of the authors
\cite{Str02}, we investigate refined enumerations of some symmetry classes
of the ASMs. In section 2 we reproduce the necessary results of paper
\cite{Str02}. In section 3 the refined enumerations for ASMs with U-turn
boundary and vertically symmetric ASMs are found, see Corollary \ref{c}.
In section 4 we prove that the refined enumerations of vertically symmetric
ASMs and off-diagonally symmetric ASMs coincide. This is the conjecture by
Kutin and Yuen presented in a message~\cite{Ku} sent by Kuperberg to the
Domino forum. In the next two paragraphs we actually quote that message.

The conjecture concerns the vertically symmetric ASMs (VSASMs)
and the off-diagonally symmetric ASMs (OSASMs). These symmetry classes of
ASMs are defined in the following way. An ASM is said to be vertically
symmetric if it remains unchanged after the reflection with respect to the
vertical line which divides the matrix into two equal parts. Note that an
ASMs of odd order only can be vertically symmetric. An ASM is said to be
off-diagonally symmetric it remains unchanged after the reflection
with respect to its diagonal, and whose diagonal is null. In paper
\cite{Kup02} G.~Kuperberg proved that there are the same number of $2n
\times 2n$ OSASMs as $(2n+1) \times (2n+1)$ VSASMs. To check this
equinumeration Kutin and Yuen did a computer experiment, and found
that there are still the same number if one fixes the position of the 1 in
the right-most column. Note that any ASM has only one 1 in the right-most
column. Here one has to remind that a VSASM cannot have the right-most 1 at
the top or bottom, and OSASMs cannot have the right-most 1 at the bottom,
so for both classes there are $2n-1$ available positions.

For example, compare the VSASMs of order~5, Figure
\ref{f:VSAMS},\footnote{As is now customary for alternating-sign matrices,
we write $+$ instead of 1 and $-$ instead of $-1$.}
\begin{figure}[ht]
\[
\left( \begin{array}{ccccc}
0 & 0 & + & 0 & 0 \\
+ & 0 & - & 0 & + \\
0 & 0 & + & 0 & 0 \\
0 & + & - & + & 0 \\
0 & 0 & + & 0 & 0
\end{array} \right)
\quad
\left( \begin{array}{ccccc}
0 & 0 & + & 0 & 0 \\
0 & + & - & + & 0 \\
+ & - & + & - & + \\
0 & + & - & + & 0 \\
0& 0 &+& 0& 0
\end{array} \right)
\quad
\left( \begin{array}{ccccc}
0 & 0 & + & 0 & 0 \\
0 & + & - & + & 0 \\
0 & 0 & + & 0 & 0 \\
+ & 0 & - & 0 & + \\
0 & 0 & + & 0 & 0
\end{array} \right),
\]
\caption{VSASMs of order 5}
\label{f:VSAMS}
\end{figure}
with the \hbox{OSASMs} of \hbox{order}~4, Figure \ref{OSAMS}.
\begin{figure}[ht]
\[
\left( \begin{array}{cccc}
0 & 0 & 0 & + \\
0 & 0 & + & 0 \\
0 & + & 0 & 0 \\
+ & 0 & 0 & 0
\end{array} \right)
\quad
\left( \begin{array}{cccc}
0 & 0 & + & 0 \\
0 & 0 & 0 & + \\
+ & 0 & 0 & 0 \\
0 & + & 0 & 0
\end{array} \right)
\quad
\left( \begin{array}{cccc}
0 & + & 0 & 0 \\
+ & 0 & 0 & 0 \\
0 & 0 & 0 & + \\
0 & 0 & + & 0
\end{array} \right).
\]
\caption{OSASMs of order 4}
\label{OSAMS}
\end{figure}
In both cases the Kutin--Yuen numbers are 1, 1, 1.

\begin{figure}[ht]
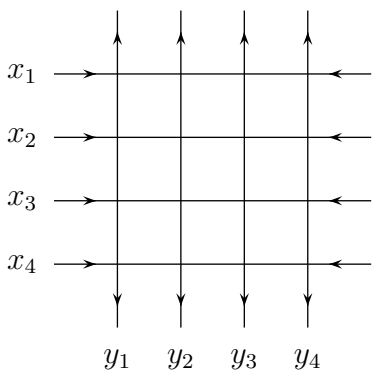

\[
\psset{unit=2em}
\pspicture[.4](-1,-.7)(5,5)
  \arrowLine(0,1)(1,1)
  \arrowLine(0,2)(1,2)
  \arrowLine(0,3)(1,3)
  \arrowLine(0,4)(1,4)
  \arrowLine(5,1)(4,1)
  \arrowLine(5,2)(4,2)
  \arrowLine(5,3)(4,3)
  \arrowLine(5,4)(4,4)
  \arrowLine(1,1)(1,0)
  \arrowLine(2,1)(2,0)
  \arrowLine(3,1)(3,0)
  \arrowLine(4,1)(4,0)
  \arrowLine(1,4)(1,5)
  \arrowLine(2,4)(2,5)
  \arrowLine(3,4)(3,5)
  \arrowLine(4,4)(4,5)
  \psline(1,1)(4,1)
  \psline(1,2)(4,2)
  \psline(1,3)(4,3)
  \psline(1,4)(4,4)
  \psline(1,1)(1,4)
  \psline(2,1)(2,4)
  \psline(3,1)(3,4)
  \psline(4,1)(4,4)
  \rput(-.5,4){$x_1$}
  \rput(-.5,3){$x_2$}
  \rput(-.5,2){$x_3$}
  \rput(-.5,1){$x_4$}
  \rput(1,-.5){$y_1$}
  \rput(2,-.5){$y_2$}
  \rput(3,-.5){$y_3$}
  \rput(4,-.5){$y_4$}
\endpspicture
\]
\caption{Square ice with domain wall boundary}
\label{f:dw}
\end{figure}

Our consideration is based on the bijection between the states of the
square ice model with appropriate boundary conditions and ASMs. Consider a
subset of vertices and edges of a square grid, such that each vertex is
either tetravalent or univalent. A state of a corresponding square ice
model is an orientation of the edges, such that two edges enter and leave
every tetravalent vertex. If orientations of the edges belonging to
univalent vertices is fixed we said that a boundary condition for the
square ice model is given. The case of importance for studying the ASMs is
the square ice model with the domain wall boundary conditions. A pattern
for a state is given in Figure \ref{f:dw}.
The meaning of the labels $x_i$ and $y_i$ will be explained later. Such
boundary conditions were first considered by Korepin \cite{Kor82}.

\begin{figure}[ht]
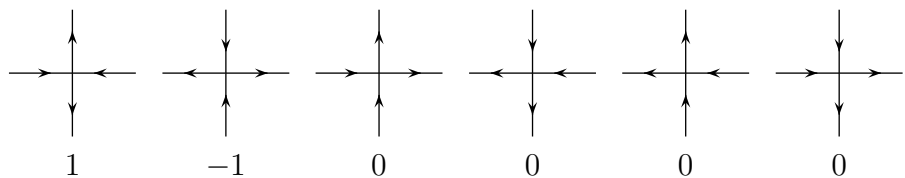

\[
\psset{unit=2em}
\begin{array}{cccccc}
\vertexA & \vertexB & \vertexC & \vertexD & \vertexE & \vertexF \\
1 & -1 & 0 & 0 & 0 & 0
\end{array}
\]
\caption{The correspondence between the square ice vertices
and the entries of alternating-sign matrices}
\label{f:dwasm}
\end{figure}

\begin{figure}[ht]
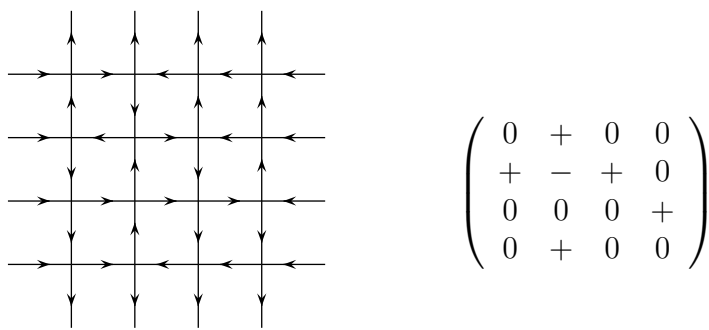

\[
\psset{unit=2em}
\pspicture[.4](0,0)(5,5)
  \arrowLine(0,4)(1,4)
  \arrowLine(1,4)(2,4)
  \arrowLine(3,4)(2,4)
  \arrowLine(4,4)(3,4)
  \arrowLine(5,4)(4,4)
  \arrowLine(0,3)(1,3)
  \arrowLine(2,3)(1,3)
  \arrowLine(2,3)(3,3)
  \arrowLine(4,3)(3,3)
  \arrowLine(5,3)(4,3)
  \arrowLine(0,2)(1,2)
  \arrowLine(1,2)(2,2)
  \arrowLine(2,2)(3,2)
  \arrowLine(3,2)(4,2)
  \arrowLine(5,2)(4,2)
  \arrowLine(0,1)(1,1)
  \arrowLine(1,1)(2,1)
  \arrowLine(3,1)(2,1)
  \arrowLine(4,1)(3,1)
  \arrowLine(5,1)(4,1)
  \arrowLine(1,4)(1,5)
  \arrowLine(1,3)(1,4)
  \arrowLine(1,3)(1,2)
  \arrowLine(1,2)(1,1)
  \arrowLine(1,1)(1,0)
  \arrowLine(2,4)(2,5)
  \arrowLine(2,4)(2,3)
  \arrowLine(2,2)(2,3)
  \arrowLine(2,1)(2,2)
  \arrowLine(2,1)(2,0)
  \arrowLine(3,4)(3,5)
  \arrowLine(3,3)(3,4)
  \arrowLine(3,3)(3,2)
  \arrowLine(3,2)(3,1)
  \arrowLine(3,1)(3,0)
  \arrowLine(4,4)(4,5)
  \arrowLine(4,3)(4,4)
  \arrowLine(4,2)(4,3)
  \arrowLine(4,2)(4,1)
  \arrowLine(4,1)(4,0)
\endpspicture
\qquad \qquad
\left( \begin{array}{cccc}
0 & + & 0 & 0 \\
+ & - & + & 0 \\
0 & 0 & 0 & + \\
0 & + & 0 & 0
\end{array} \right)
\]
\caption{An example of the correpondence between the square ice states and
alternating-sign matrices}
\label{f:dwasmex}
\end{figure}

If we replace each tetravalent vertex of a state of the square ice model
with the domain wall boundary condition by a number according Figure
\ref{f:dwasm} we will obtain a matrix.
An example is given in Figure \ref{f:dwasmex}.
It is not difficult to check that in this way we come to the bijection
between the states and the ASMs \cite{EKLP92}.

\begin{figure}[ht]
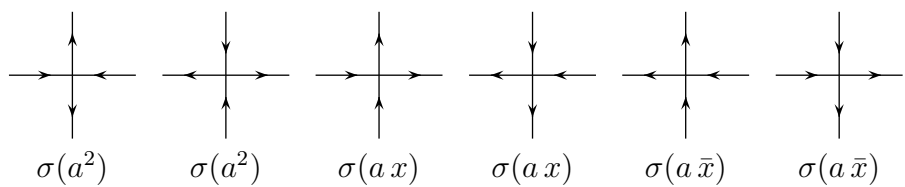

\[
\psset{unit=2em}
\begin{array}{cccccc}
\vertexA & \vertexB & \vertexC & \vertexD & \vertexE & \vertexF \\
\sigma(a^2) & \sigma(a^2) & \sigma(a \, x) & \sigma(a \, x) & \sigma(a \,
\bar x) & \sigma(a \, \bar x)
\end{array}
\]
\caption{The weights of the vertices with the spectral parameter $x$}
\label{f:wghts}
\end{figure}

The above correspondence was first used to solve ASMs enumeration problems
by Kuperberg \cite{Kup96}. His strategy was as follows. Consider the
partition function of the square ice model with the domain wall boundary
conditions. This is the sum of the weights of all possible states. The
weight of a state is the product of the weights of all tetravalent vertices
which are defined in the following way. Associate spectral parameters $x_i$
with the vertical lines of the grid and spectral parameters $y_i$ with the
horizontal ones, see Figure \ref{f:dw}. A vertex at the intersection of the
line with the spectral parameters $x_i$ and $y_j$ is supplied with the
spectral parameter $x_i/y_j$. The weights of the vertices depend on the
value of the spectral parameter and are given in
Figure~\ref{f:wghts},\footnote{In contrast with paper \cite{Str02} we use
multiplicative parameters.}
where $a$ is a parameter common for all vertices and we use the convenient
abbreviations
\begin{eqnarray*}
& \bar x = x^{-1}, \\
& \sigma(x) = x - \bar x
\end{eqnarray*}
introduced by Kuperberg.

Denote the partition function of the square ice model with the domain wall
boundary condition by $Z(n; \bm x, \bm y)$. Here $n$ is the size of the
square ice, $\bm x$ and $\bm y$ are row-vectors constructed from the
spectral parameters,
\[
\bm x = (x_1, \ldots, x_n), \qquad \bm y = (x_1, \ldots, x_n).
\]
Given a number $x$, denote by $A(n; x)$ the total weight of the $n \times
n$ ASMs, where the weight of an individual ASM is $x^k$ if it has $k$
matrix elements equal to $-1$. Consider the partition function $Z(n; \bm x,
\bm y)$
for $\bm x = \bm 1$ and $\bm y = \bm 1$, where
\[
\bm 1 = (1, 1, \ldots, 1).
\]
If the number of $-1$'s in an $n \times n$ alternating-sign matrix is
equal to $k$, then the number of 1's is equal to $n+k$ and the number of
0's is $n^2-n-2k$. It is clear that the weight of the corresponding state
of the square ice is
\[
\sigma(a^2)^k \sigma(a^2)^{n+k} \sigma(a)^{n^2-n-2k} = \sigma(a)^{n^2-n}
\sigma(a^2)^n \left[ \frac{\sigma(a^2)}{\sigma(a)} \right]^{2k}.
\]
Hence we have the equality
\begin{equation}
A(n; x) = \frac{1}{\sigma(a)^{n^2-n} \sigma(a^2)^n} \, Z(n; \bm 1, \bm 1),
\label{1}
\end{equation}
where
\[
x = \left[ \displaystyle \frac{\sigma(a^2)}{\sigma(a)} \right]^2 = (a +
\bar a)^2.
\]
In particular, when $a = \exp(\rmi \, \pi/3)$ we have $x=1$, and relation
(\ref{1}) gives the total number of $n \times n$ ASMs
\[
A(n) = \left. \frac{1}{\sigma(a)^{n^2}} \, Z(n; \bm 1, \bm 1) \right|_{a =
\exp(\rmi \, \pi/3)}.
\]
Thus, the enumeration problem for ASMs is reduced to the problem of
studying
the partition function $Z(n; \bm x, \bm y)$ for $\bm x = \bm 1$ and $\bm y
= \bm 1$.

In paper \cite{Kup96} Kuperberg used for $Z(n; \bm x, \bm y)$ the
representation via the Izergin--Korepin determinant \cite{Ize87, KIB93} and
proved the formula for $A(n)$ conjectured by Mills, Robbins and Rumsey
\cite{MRR82, MRR83} and first proved by Zeilberger \cite{Zei96}. In paper
\cite{Kup02} Kuperberg generalised the Izergin--Korepin determinant and the
Tsuchiya determinant \cite{Tsu} to address enumeration problems related to
numerous symmetry classes of alternating-sign matrices.

In paper \cite{Zei96a} Zeilberger used again the representation of the
partition function $Z(n, \bm x, \bm y)$ via the Izergin--Korepin
determinant to prove the so-called refined ASM conjecture by Mills, Robbins
and Rumsey \cite{MRR82, MRR83}. Let us explain what this enumeration means.
Note that the first column of an ASM contains only one 1, all other entries
are 0's. Therefore, we can try to enumerate the ASMs for which the unique 1
is at the $r$th position in the first column. The corresponding numbers
$A(n, r)$ give the refined enumeration of the ASMs. Similarly to the case
of the usual enumeration one can assume that the weight of an individual
ASM is $x^k$ if it has $k$ matrix elements equal to $-1$. This gives the
polynomials $A(n, r; x)$ which describe the weighted refined enumeration of
the ASMs.

To relate the polynomials $A(n, r; x)$ to the partition function $Z(n; \bm
x, \bm y)$ consider the case where $\bm x = \bm 1$ and $\bm y = (u, 1,
\ldots, 1)$. In this case the label of a vertex which belongs to the first
column is $\bar u$, otherwise it is $1$. Consider a state of the square ice
for which the unique 1 in the first column of the corresponding ASM belongs
to the $r$th row. Here in the first column we have one vertex of
the first type, $r-1$ vertices of the third type and $n-r$ vertices of the
sixth type, where type is defined in accordance with the position of the
vertex in Figure \ref{f:wghts}. Hence, the contribution of the first column
to the weight of the state is $\sigma(a^2) \, \sigma(a \, \bar u)^{r-1} \,
\sigma(a \, u)^{n-r}$. Denote by $k$ the number of $-1$'s in the ASM under
consideration. Then the number of 1's in the ASM with the first column
removed is equal to $n -1 + k$, and the number of 0's is equal to $n^2 -
2n + 1 - 2k$. Thus, the weight of the state under consideration is
\[
\sigma(a)^{n^2-2n+1} \, \sigma(a^2)^n \, \sigma(a \, u)^{n-1} \left[
\frac{\sigma(a^2)}{\sigma(a)} \right]^{2k} \left[
\frac{\sigma(a \, \bar u)}{\sigma(a \, u)} \right]^{r-1},
\]
and we have the equality
\begin{equation}
\sum_{r=1}^n A(n, r; x) \, t^{r-1} = \frac{Z(n; \bm 1, (u,1,\ldots,1))}
{\sigma(a)^{n^2-2n+1} \, \sigma(a^2)^n \, \sigma(a \, u)^{n-1}},
\label{6}
\end{equation}
where
\[
x = \left[ \displaystyle \frac{\sigma(a^2)}{\sigma(a)} \right]^2, \qquad
t = \frac{\sigma(a \, \bar u)}{\sigma(a \, u)}.
\]
Thus, the refined enumeration problem for ASMs is reduced to the problem of
studing the partition function $Z(n; \bm x, \bm y)$ for $\bm x = \bm 1$ and
$\bm y = (u, 1, \ldots, 1)$. 

Using the representaion of the partition function via the Izergin--Korepin
determinat, Kuperberg and Zeilberger had to overcome its singularity at
desired values of the spectral parameters. One of the authors of the
present paper proposed a new way to deal with the Izergin--Korepin
determinat valid at $a = \exp(\rmi \, \pi/3)$. It allowed to give
more simple proof of the refined ASM conjecture. In the present paper we
use the method of paper \cite{Str02} to obtain formulas for the refined
enumeration of some symmetry classes of the ASMs and prove the Kutin--Yuen
conjecture.

\section{The case of general ASMs}

In this section we describe the method to investigate refined enumerations
proposed in paper \cite{Str02}. We obtain at first a new determinat
representation for the partition function $Z(n; \bm x, \bm y)$ valid for $a
= \exp(\rmi \, \pi/3)$. Actually this representaion is not connected
directly with enumeration problems, but reveals a usefull symmetry of the
partition function. 

Let us start with the representation of the partition function $Z(n; \bm x,
\bm y)$ via the Izergin--Korepin determinant. Recall that the proof of the
validity of that representation is based on the following three lemmas.

\begin{lemma} \label{l:1}
The partition function $Z(n; \bm x, \bm y)$ is symmetric in the coordinates
of $\bm x$ and in the coordinates of $\bm y$.
\end{lemma}

\begin{lemma} \label{l:2}
The partition function $Z(n; \bm x, \bm y)$ satisfies the recurrent
relation
\[
\frac{Z(n; (x_1, \ldots, x_{n-1}, x_n), (y_1, \ldots, y_{n-1} , a \,
x_n))}{Z(n-1; (x_1, \ldots, x_{n-1}), (y_1, \ldots, y_{n-1}))} =
\sigma(a^2) \prod_{k = 1}^{n-1} \sigma(a \, \bar x_n \, y_k) \prod_{k =
1}^{n-1} \sigma(a \, \bar x_k \, y_n).
\]
\end{lemma}

\begin{lemma} \label{l:3}
For each $i = 1, \ldots, n$ the product $x_i^{n-1} Z(n; \bm x, \bm y)$ is a
polynomial in $x_i^2$ of degree $n-1$. For each $i = 1, \ldots, n$ the
product $y_i^{n-1} Z(n; \bm x, \bm y)$ is a polynomial in $y_i^2$ of
degree $n-1$.
\end{lemma}

\noindent It is clear that these three lemmas, together with the initial
condition $Z(1, \bm x, \bm y) = \sigma(a^2)$, inductively define $Z(n, \bm
x, \bm y)$. The representation of $Z(n, \bm x, \bm y)$  via the
Izergin--Korepin determinant is the content of the following theorem.

\begin{theorem}
The partition function $Z(n; \bm x, \bm y)$ can be represented as
\begin{equation}
Z(n; \bm x, \bm y) = \frac{\displaystyle \sigma(a^2)^n \prod_{i,j}
\alpha(x_i \, \bar y_j)}{\displaystyle \prod_{i < j} \sigma(\bar x_i \,
x_j) \, \sigma(y_i \, \bar y_j)} \, \det M(n; \bm x, \bm y), \label{2}
\end{equation}
where $M(n; \bm x, \bm y)$ is an $n \times n$ matrix defined by the
equality
\[
M(n; \bm x, \bm y)_{ij} = \frac{1}{\alpha(x_i \, \bar y_j)}.
\]
and the function $\alpha$ is defined by
\[
\alpha(x) = \sigma(a \, x) \, \sigma(a \, \bar x).
\]
\end{theorem}

\noindent One can prove this theorem directly checking that the right side
of equality (\ref{2}) satisfies Lemmas \ref{l:1}, \ref{l:2} and \ref{l:3}.

Consider the case of $a = \exp(\rmi \, \pi/3)$. Note that in this case one
has the equalities
\begin{equation}
1 - a + a^2 = 0, \qquad 1 - \frac{1}{a} + \frac{1}{a^2} = 0,
\label{3}
\end{equation}
which can be used to prove that
\[
\sigma(a^2) = \sigma(a)
\]
and that
\[
\sigma(x) \, \sigma(ax) \, \sigma(a^2 x) = - \sigma(x^3).
\]
The last equality implies
\begin{equation}
\alpha(x) = - \frac{\sigma(x^3)}{\sigma(x)}. \label{8}
\end{equation}
Using this formula and denoting denote $u_{2i-1} = x_i$, $u_{2i} = y_i$, we
rewrite equality (\ref{2}) as
\[
Z(n; \bm u) = \frac{\displaystyle (-1)^n \sigma(a)^n \prod_{i,j}
\sigma(u_{2i-1}^3 \, \bar u_{2j}^3)}{\displaystyle \prod_{i
< j} [\sigma(\bar u_{2i-1} \, u_{2j-1}) \, \sigma(u_{2i} \, \bar u_{2j})]
\prod_{i,j} \sigma(u_{2i-1} \, \bar u_{2j})} \, \det M(n; \bm u).
\]
Introduce now the function
\begin{equation}
F(n; \bm u) = \prod_{\mu < \nu} \sigma(u_\mu \, \bar u_\nu)
\, Z(n; \bm u). \label{5}
\end{equation}
Here and below we assume that Greek indices run from 1 to $2n$. Using the
determinant representation for $Z(n; \bm u)$, we can also write
\begin{equation}
F(n; \bm u) = (-1)^n \sigma(a)^{n} \prod_{i,j} \sigma(u_i^3 \, \bar
u_{i+n}^3) \det M(n; \bm u). \label{4}
\end{equation}
The following simple lemma is very important for our consideration.

\begin{lemma} \label{l:4}
For every $\mu = 1, \ldots, 2n$ one has
\begin{eqnarray*}
\lefteqn{F(n; (u_1, \ldots, u_\mu, \ldots, u_{2n}))} \\[.5em]
&& {} + F(n; (u_1, \ldots, a^2 u_\mu, \ldots, u_{2n})) + F(n; (u_1,
\ldots, a^4 u_\mu, \ldots, u_{2n})) = 0.
\end{eqnarray*}
\end{lemma}

\begin{proof}
For $a = \exp(\rmi \, \pi/3)$ one has
\[
M(n, \bm x, \bm y)_{ij} = - \frac{\sigma(x_i \, \bar y_j)}{\sigma(x_i^3 \,
\bar y_j^3)}.
\]
Using equalities (\ref{3}), we obtain
\begin{equation}
\sigma(x) + \sigma(a^2 x) + \sigma(a^4 x) = 0 \label{19}
\end{equation}
that gives for the first column
\[
M(n; (x_1, \ldots, x_n), \bm y)_{1j} + M(n; (a^2 x_1, \ldots, x_n), \bm
y)_{1j} + M(n; (a^4 x_1, \ldots, x_n), \bm y)_{1j} = 0.
\]
Since $\det M(n; \bm x, \bm y)$ for every $j$ linearly depends on $M(n; \bm
x, \bm y)_{1j}$, the above equality implies that
\begin{eqnarray*}
\lefteqn{\det M(n; (x_1, \ldots, x_n), \bm y)} \\[.5em]
&& {} + \det M(n; (a^2 x_1, \ldots, x_n), \bm y) + \det M(n; (a^4 x_1,
\ldots, x_n), \bm y) = 0.
\end{eqnarray*}
Taking into account equality (\ref{4}), we conclude that the statement of
the lemma is valid for $\mu=1$. For all other values of $\mu$ one can
apply the same reasonings.
\end{proof}

\noindent Lemma \ref{l:3} and relation (\ref{5}) imply the following
lemma.

\begin{lemma} \label{l:5}
For every $\mu = 1, \ldots, 2n$ the function $u_\mu^{3n-2} F(n, \bm u)$ is
a polynomial of degree $3n-2$ in $u_\mu^2$.
\end{lemma}

\noindent We also have the following evident lemma.

\begin{lemma} \label{l:6}
The function $F(n, \bm u)$ turns to zero if $u_\mu^2 = u_\nu^2$ for some
$\mu \ne \nu$.
\end{lemma}

Let us show now that Lemmas \ref{l:4}, \ref{l:5} and \ref{l:6} determine
the function $F(n, \bm u)$ uniquely up to a constant factor.

\begin{lemma} \label{l:7}
The function $F(n, \bm u)$ satisfying Lemmas \ref{l:4}, \ref{l:5} and
\ref{l:6} is proportional to the determinant of the matrix
\[
P(n; \bm u) = \left( \begin{array}{ccccc}
u_1^{3n-2} & u_2^{3n-2} & u_3^{3n-2} & \ldots & u_{2n}^{3n-2} \\[.4em]
u_1^{3n-4} & u_2^{3n-4} & u_3^{3n-4} & \ldots & u_{2n}^{3n-4} \\[.4em]
u_1^{3n-8} & u_2^{3n-8} & u_3^{3n-8} & \ldots & u_{2n}^{3n-8} \\
\vdots & \vdots & \vdots & \ddots & \vdots \\
u_1^{-3n+2} & u_2^{-3n+2} & u_3^{-3n+2} & \ldots & u_{2n}^{-3n+2} \\
\end{array} \right).
\]
\end{lemma}

\begin{proof}
Taking into account Lemma \ref{l:5}, for some fixed value of $\mu$ we
write
\[
F(n; \bm u) = \sum_{k=1}^{3n - 1} a_k^{(\mu)}(u_1, \ldots, \hat
u_\mu, \ldots, u_{2n}) \, u^{3n - 2k}_\mu,
\]
where hat means that the corresponding argument is omitted. From Lemma
\ref{l:4} it follows that $a_k^{(\mu)} = 0$ if $k$ is divisible by 3.
Therefore, one has
\[
F(n; \bm u) = \sum_{\scriptstyle k=1 \atop \scriptstyle k \ne 0 \ \mathrm{
mod} \ 3}^{3n - 1} a_k^{(\mu)}(u_1, \ldots, \hat u_\mu, \ldots, u_{2n}) \,
u^{3n - 2k}_\mu.
\]
Hence, we have only $2n$ unknown functions $a_k^{(\mu)}$. Lemma
\ref{l:6} implies that for any $\mu \ne \nu$ one has
\[
\sum_{\scriptstyle k=1 \atop \scriptstyle k \ne 0 \ \mathrm{
mod} \ 3}^{3n - 1} a_k^{(\mu)}(u_1, \ldots, \hat u_\mu, \ldots, u_{2n})
\, u^{3n - 2k}_\nu = 0.
\]
Thus, we have a system of $2n - 1$ linear equations for $2n$ functions
$a_k^{(\mu)}$. In general position the rank of this system is equal to
$2n - 1$. Hence, it determines the functions $a_k^{(\mu)}$ uniquely up
to a common factor which cannot depend on $u_\mu$. Considering all
values of $\mu$, we conclude that Lemmas \ref{l:4}, \ref{l:5} and
\ref{l:6} determine the function $F(n, \bm u)$ uniquely up to a constant
factor. It is easy to see that the determinant of the matrix $P(n, \bm u)$
satisfies Lemmas \ref{l:4}, \ref{l:5} and \ref{l:6}. Hence, the function
$F(n; \bm u)$ must be proportional to this determinant.
\end{proof}

\begin{lemma} \label{l:6a}
The determinant of the matrix $P(n; \bm u)$ satisfies the recurrent
relation
\[
\frac{\det P(n; (u_1, \ldots, u_{2n-2}, u_{2n-1}, a \, u_{2n-1}))}{\det
P(n-1; (u_1, \ldots, u_{2n-2}))} = (-1)^n \sigma(a) \prod_{\mu=1}^{2n-2}
\sigma(u_\mu^3 \, \bar u_{2n-1}^3).
\]
\end{lemma}

\begin{proof}
Recalling that we consider the case of $a = \exp(\rmi \, \pi/3)$, one can
easily see that the determinant of the matrix $P(n; \bm u)$ for $u_{2n} = a
\, u_{2n-1}$ is
\[
(-1)^{n+1} \left| \begin{array}{ccccc}
u_1^{3n-2} & u_2^{3n-2} & \ldots & u_{2n-1}^{3n-2} & a \, u_{2n-1}^{3n-2}
\\[.4em]
u_1^{3n-4} & u_2^{3n-4} & \ldots & u_{2n-1}^{3n-4} & \bar a \,
u_{2n-1}^{3n-4} \\[.4em]
u_1^{3n-8} & u_2^{3n-8} & \ldots & u_{2n-1}^{3n-8} & a \, u_{2n-1}^{3n-8}
\\
\vdots & \vdots & \ddots & \vdots & \vdots \\
u_1^{-3n+8} & u_2^{-3n+8} & \ldots & u_{2n-1}^{-3n+8} & \bar a \,
u_{2n-1}^{-3n+8} \\[.4em]
u_1^{-3n+4} & u_2^{-3n+4} & \ldots & u_{2n-1}^{-3n+4} & a \,
u_{2n-1}^{-3n+4} \\[.4em]
u_1^{-3n+2} & u_2^{-3n+2} & \ldots & u_{2n-1}^{-3n+2} & \bar a \,
u_{2n-1}^{-3n+2} \\
\end{array} \right|.
\]
Let us subsrtact from the first row of the above determinant its third row
multiplied by $u_{2n-1}^6$, subsrtact from the second row the forth row
again multiplied by $u_{2n-1}^6$, and so on whenever is possible. As the
result we have the following expression for $\det P(n; \bm u)$:
\[
(-1)^{n+1} \left| \begin{array}{ccccc}
(u_1^6 - u_{2n-1}^6) \, u_1^{3n-8} & (u_2^6 - u_{2n-1}^6) \, u_2^{3n-8} &
\ldots & 0 & 0 \\[.4em]
(u_1^6 - u_{2n-1}^6) \, u_1^{3n-10} & (u_2^6 - u_{2n-1}^6) \, u_2^{3n-10} &
\ldots & 0 & 0 \\[.4em]
(u_1^6 - u_{2n-1}^6) \, u_1^{3n-14} & (u_2^6 - u_{2n-1}^6) \, u_2^{3n-14} &
\ldots & 0 & 0 \\
\vdots & \vdots & \ddots & \vdots & \vdots \\
(u_1^6 - u_{2n-1}^6) \, u_1^{-3n+2} & (u_2^6 - u_{2n-1}^6) \, u_2^{3n+2} &
\ldots & 0 & 0 \\
u_1^{-3n+4} & u_2^{-3n+4} & \ldots & u_{2n-1}^{-3n+4} & a \,
u_{2n-1}^{-3n+4} \\[.4em]
u_1^{-3n+2} & u_2^{-3n+2} & \ldots & u_{2n-1}^{-3n+2} & \bar a \,
u_{2n-1}^{-3n+2} \\[.4em]
\end{array} \right|.
\]
Now the statement of the Lemma is evident.
\end{proof}

\begin{theorem} \label{t:2}
For $a = \exp(\rmi \, \pi/3)$ the partition function $Z(n; \bm u)$ has the
following determinant representation:
\begin{equation}
Z(n; \bm u) = (-1)^{\frac{n(n-1)}{2}} \frac{\sigma(a)^n}{\displaystyle
\prod_{\mu < \nu} \sigma(u_\mu \, \bar u_\nu)} \, \det P(n; \bm
u). \label{24}
\end{equation}
\end{theorem}

\begin{proof} One can show that in our case the recurrent relation of Lemma
\ref{l:2} takes the form
\[
\frac{Z(n; (u_1, \ldots, u_{2n-2}, u_{2n-1}, a \, u_{2n-1}))}{Z(n-1; (u_1,
\ldots u_{2n-2}))} = \sigma(a) \prod_{\mu=1}^{2n-2} \sigma(a \, u_\mu \,
\bar u_{2n-1}).
\]
Taking this fact and Lemma \ref{l:6a} into account, we easily see that the
right hand side of the relation (\ref{24}) satisfies Lemmas \ref{l:1},
\ref{l:2} and \ref{l:3}. Therefore, since the equality (\ref{24}) is valid
for $n=1$, it is valid for any $n$.
\end{proof}

\noindent Theorem \ref{t:2} shows, in particular, that the partition
function $Z(n; \bm u)$ is symmetric in the coordinates of the vector $\bm
u$. Returning to the vectors $\bm x$ and $\bm y$, we can formulate the
following corollary, which generalises Lemma \ref{l:1}.

\begin{corollary}
For $a = \exp(\rmi \, \pi/3)$ the partition function $Z(n, \bm x, \bm y)$
is symmetric in the union of the coordinates of $\bm x$ and $\bm y$.
\end{corollary}

Let us return to enumeration problems. As was noted in section \ref{s:1},
to solve them we should consider the partition function $Z(n; \bm x, \bm
y)$ for $\bm x = \bm 1$, $\bm y = 1$, or for $\bm x = 1$, $\bm y = (u, 1,
\ldots, 1)$. Unfortunately, both determinants, $\det M(n, \bm u)$ and $\det
P(n, \bm u)$, become singular under such specialisation of the spectral
parameters. Kuperberg \cite{Kup96} proposed to find values of the
Izerging--Korepin determinant along a curve that includes a desired values
of the spectral parameters. The same method was used by Zeibelberg
\cite{Zei96a} to prove the refined ASM conjecture. The method used in paper
\cite{Str02} is different. To follow this method let us define for a fixed
value of $\mu$ the function
\[
F^{(\mu)}(n, \bm u) = \prod_{\nu \ne \mu} \sigma(u_\mu \, \bar
u_\nu) Z(n, \bm u),
\]
and specialise then the values of the spectral parameters in the following
way
\[
f(n, u) = F^{(\mu)}(n, (\, \underbrace{1, \ldots, 1,}_{\mu-1} u,
\underbrace{1, \ldots , 1}_{2n - \mu} \,)).
\]
Since the partition function $Z(n; \bm u)$ is symmetric in the coordinates
of $\bm u$, the function $f(n; u)$ does not depend on $\mu$. Actually,
one can write
\[
f(n; u) = \sigma(u)^{2n-1} Z(n, (u, 1, \ldots, 1)).
\]

\begin{lemma} \label{l:8}
The function $f(n; u)$ has the following properties.
\begin{itemize}

\item[\rm (a)] The function $u^{3n-2} f(n; u)$ is a polynomial of degree
$3n-2$ in $u^2$.

\item[\rm (b)] The function $f(n; u)$ satisfies the relation
\[ f(n; \bar u) = - f(n; u).
\]

\item[\rm (c)] The function $f(n; u)$ obeys the equality
\[
f(n; u) + f(n; a^2 u) + f(n; a^4 u) = 0.
\]

\item[\rm (d)] The Laurent polynomial $f(n; u)$ is divisible by
$\sigma(u)^{2n-1}$.

\end{itemize}
\end{lemma}
Probably, only the property (b) has to be discussed. Actually it follows
from the equality
\[
Z(n; (\bar u_1, \bar u_2, \ldots, \bar u_{2n})) = Z(n; (u_1, u_2, \ldots,
u_n))
\]
which is a consequence of the left-right and top-bottom symmetries of the
square ice with the domain wall boundary.

The next important lemma was proved by one of the authors in paper
\cite{Str01}.

\begin{lemma} \label{l:9}
A function $f(n; u)$ satisfying Lemma \ref{l:8} is proportional to the
function
\[
\varphi(n, u) = \frac{(-1)^{n-1}}{ \sigma(a) {2n-2 \choose
n-1}} \sum_{k=0}^{n-1} {n-\frac{4}{3} \choose n-1-k} {n-\frac{2}{3}
\choose k} \sigma(u^{3n - 2 - 6k}).
\]
The function $\varphi(n; u)$ is normalized by the condition
\begin{equation}
\varphi(n; a) = 1.
\label{7}
\end{equation}
\end{lemma}
The normalization condition (\ref{7}) can be verified using the identity
\[
\sum_{k=0}^{n-1} {n-\frac{4}{3} \choose n-1-k} {n-\frac{2}{3}
\choose k} = {2n - 2 \choose n - 1}.
\]

Equality (\ref{6}) for $a = \exp(\rm i \, \pi/3)$ and $u = a$ gives
\[
Z(n, (a, 1, \ldots, 1) = A(n-1) \, \sigma(a)^{n^2},
\]
where we have taken into account the evident equality $A(n, 1) = A(n-1)$.
Therefore, we have
\[
f(n, u) = A(n-1) \, \sigma(a)^{n^2+2n-1} \varphi(n; u)
\]
that implies
\[
Z(n, (u, 1, \ldots, 1) = A(n-1) \, \sigma(a)^{n^2+2n-1} \frac{\varphi(n;
u)}{\sigma(u)^{2n-1}}.
\]
It is instructive to rewrite equality (\ref{6}) for $a = \exp(\rmi \,
\pi/3)$ in terms of the function $\varphi(n; u)$:
\begin{equation}
\frac{1}{A(n-1)} \sum_{r=1}^n A(n,r) \, t^{r-1} = \frac{\sigma(a)^{3n-2} \,
\varphi(n; u)}{\sigma(au)^{n-1} \, \sigma(u)^{2n-1}}. \label{11}
\end{equation}
In paper \cite{Str02} this relation was used to prove the refined ASM
conjecture and to obtain some results on doubly refined ASM enumeration. In
the next sections we will show that the method described above
effectively works in other cases.

\section{ASMs with U-turn boundary}

The Kutin--Yuen conjecture concerns vertically symmetric and off-diagonally
symmetric alter\-nating-sign matrices. One can formulate the ice square
model related to vertically symmetric ASMs, but as far as we know there is
no a determinant formula for this model with arbitrary spectral parameters.
However, one can show that the set of vertically symmetric matrices can be
considered as a subset of the so-called U-turn alternating-sign matrices
(UASMs). Let us define consider these matrices and the partition function
for the corresponding square ice model.

\begin{figure}[ht]
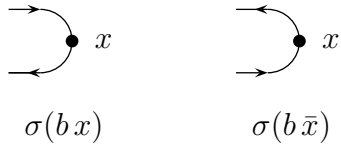

\[
\psset{unit=2em}
\begin{array}{ccc}
  \uvertexRight & \qquad   & \uvertexLeft \\[1.5em]
  \sigma(b \, x) & \qquad  &   \sigma(b \, \bar x)
\end{array}
\]
\caption{Weights of U-turn vertices}
\label{f:uwghts}
\end{figure}

Recall \cite{Kup02} that a $2n \times n$ ASM with U-turn boundary looks
vertically as a usual ASM, horizontally the 1's and $-1$'s alternate if we
walk along an odd row from left to right and then along the next even row
from right to left, see example in Figure~\ref{f:uasmex}.
The states of the corresponding square ice model are constructed in
accordance with pattern given in Figure~\ref{f:uasm}.
\begin{figure}[ht]
\centering
\begin{minipage}[b]{.425\linewidth}
\[
\left(
  \psset{unit=3em}
  \pspicture[.45](0,0)(1.8,3)
    \rput(.25,.25){$0$}
    \rput(.75,.25){$0$}
	\rput(1.25,.25){$0$}
    \rput(.25,.75){$0$}
    \rput(.75,.75){$+$}
	\rput(1.25,.75){$0$}
    \rput(.25,1.25){$0$}
    \rput(.75,1.25){$0$}
	\rput(1.25,1.25){$+$}
    \rput(.25,1.75){$+$}
    \rput(.75,1.75){$-$}
	\rput(1.25,1.75){$0$}
	\rput(.25,2.25){$0$}
	\rput(.75,2.25){$+$}
	\rput(1.25,2.25){$-$}
	\rput(.25,2.75){$0$}
	\rput(.75,2.75){$0$}
	\rput(1.25,2.75){$+$}
    \psbezier(1.5,.75)(1.8,.75)(1.8,.25)(1.5,.25)
	\psbezier(1.5,1.75)(1.8,1.75)(1.8,1.25)(1.5,1.25)
	\psbezier(1.5,2.75)(1.8,2.75)(1.8,2.25)(1.5,2.25)
  \endpspicture
\right.
\]
\caption{Example of UASM}
\label{f:uasmex}
\end{minipage}
\hspace{.05\linewidth}
\begin{minipage}[b]{.425\linewidth}
\[
\psset{unit=2em}
\pspicture(0,-.7)(5,5)
  \arrowLine(1,1)(2,1)
  \arrowLine(1,2)(2,2)
  \arrowLine(1,3)(2,3)
  \arrowLine(1,4)(2,4)
  \psline(2,1)(3.5,1)
  \psline(2,2)(3.5,2)
  \psline(2,3)(3.5,3)
  \psline(2,4)(3.5,4)
  \arrowLine(2,1)(2,0)
  \arrowLine(3,1)(3,0)
  \arrowLine(2,4)(2,5)
  \arrowLine(3,4)(3,5)
  \psline(2,1)(2,4)
  \psline(3,1)(3,4)
  \psarc(3.5,1.5){.5}{270}{90}
  \psarc(3.5,3.5){.5}{270}{90}
  \qdisk(4,1.5){.1}
  \qdisk(4,3.5){.1}
  \rput(.5,1){$\bar x_2$}
  \rput(.5,2){$x_2$}
  \rput(.5,3){$\bar x_1$}
  \rput(.5,4){$x_1$}
  \rput(4.8,1.5){$ax_2$}
  \rput(4.8,3.5){$ax_1$}
  \rput(2,-.5){$y_1$}
  \rput(3,-.5){$y_2$}
\endpspicture
\]
\caption{Square ice with U-turn boundary}
\label{f:uasm}
\end{minipage}
\end{figure}
Here at the additional bivalent U-turn vertices one of the edges points in
and the other one points out. The corresponding weights are given in Figure
\ref{f:uwghts}.
Note that an additional parameter~$b$ is introduced. It can be easily shown
that we have a bijective correspondence between the states and UASMs.

The square ice model with U-turn boundary was considered first by
Tsuchiya \cite{Tsu}. He obtained the determinant formula for the
partition function. We will use its modification which was suggested by
Kuperberg \cite{Kup02} and looks as
\[
Z_{\mathrm U}(n; \bm x, \bm y) = \frac{\displaystyle \sigma(a^2)^n \prod_i
\left[ \sigma(b \, \bar y_i) \, \sigma(a^2 x_i^2) \right] \prod_{i,j}
\left[ \alpha(x_i \, \bar y_j) \, \alpha(x_i \, y_j)\right]}{\displaystyle
\prod_{i < j} \left[\sigma(\bar x_i \, x_j) \, \sigma(y_i \, \bar y_j)
\right] \prod_{i \le j} \left[ \sigma(\bar x_i \, \bar x_j) \, \sigma(y_i
\, y_j) \right]} \det M_{\mathrm U}(n; \bm x, \bm y),
\]
where $n \times n$ matrix $M_{\mathrm U}$ is defined as
\[
M_{\mathrm U}(n,\bm x, \bm y)_{i,j} = \frac{1}{\alpha(x_i \, \bar y_j)} -
\frac{1}{\alpha(x_i \, y_j)}.
\]

Denote by $A_{\mathrm U}(2n, r; x, y)$ the total weight of the UASMs whose
unique 1 in the first column belongs to the $r$th row, each $-1$ has the
multiplicative weight $x$ and each upward oriented U-turn has the
multiplicative weight $y$. To relate these numbers to the partition
function $Z_{\mathrm U}(n; \bm x, \bm y)$ consider again the case where
$\bm x = \bm 1$ and $\bm y = (u, 1, \ldots, 1)$. Let for a state of the
square ice with U-turn boundary the corresponding UASM has 1 in the
intersection of the first column and the $r$th row. Here again in the first
column we have one vertex of the first type, $r-1$ vertices of the third
type and $2n-r$ vertices of the sixth type, where type is determined again
by Figure \ref{f:wghts}. Hence, the contribution of the first column to the
weight of the state is $\sigma(a^2) \, \sigma(a \, \bar u)^{r-1} \,
\sigma(a \, u)^{n-r}$. Let $k$ be the number of $-1$'s in the ASM under
consideration. Then the number of 1's in the ASM with the first column
removed is equal to $n -1 + k$, and the number of 0's is equal to $2n^2 -
3n +1 - 2k$. Denote by $l$ the number of upward U-turns. After all one can
see that the weight of the state under consideration is
\[
\sigma(a)^{2n^2-3n+1} \, \sigma(a^2)^n \, \sigma(a \, u)^{2n-1} \, \sigma(a
\, b)^n \left[\frac{\sigma(a^2)}{\sigma(a)} \right]^{2k} \left[
\frac{\sigma(a \, \bar u)}{\sigma(a \, u)} \right]^{r-1} \left[
\frac{\sigma(\bar a \, b)}{\sigma(a \, b)} \right]^l,
\]
and we have the equality
\begin{equation}
\sum_{r=1}^{2n} A_{\mathrm U}(2n, r; x, y) \, t^{r-1} = \frac{Z_{\mathrm
U}(n; \bm 1, (u,1,\ldots,1))} {\sigma(a)^{2n^2-3n+1} \, \sigma(a^2)^n \,
\sigma(a \, u)^{2n-1} \, \sigma(a \, b)^n}, \label{9}
\end{equation}
where
\[
x = \left[ \displaystyle \frac{\sigma(a^2)}{\sigma(a)} \right]^2, \qquad y
= \frac{\sigma(\bar a \, b)}{\sigma(a \, b)}, \qquad t = \frac{\sigma(a \,
\bar u)}{\sigma(a \, u)}.
\]

It is convenient to introduce the modified partition function
\[
Z'_{\mathrm U}(n; \bm x, \bm y) = \frac{Z_{\mathrm U}(n; \bm x, \bm
y)}{\displaystyle \prod_i \left[ \sigma(b \, \bar y_i) \, \sigma(a^2
x_i^2) \right]}.
\]
Assume that $a = \exp(\rmi \, \pi/3)$. In this case, using equality
(\ref{8}), we obtain
\[
Z'_{\mathrm U}(n; \bm u) = (-1)^n \frac{\displaystyle \sigma(a)^n
\prod_{i,j} \left[ \sigma(u_{2i-1}^3 \, \bar u_{2j}^3) \, \sigma(u_{2i-1}^3
\, u_{2j}^3) \right]}{\displaystyle \prod_{\mu < \nu} \sigma(u_\mu \,
\bar u_\nu) \prod_{\mu \le \nu} \sigma(u_\mu \, u_\nu)} \det
M_{\mathrm U}(n; \bm u),
\]
where we again use the notation $u_{2i-1} = x_i$, $u_{2i} = y_i$. Consider
the function
\[
F_{\mathrm U}(n; \bm u) = \prod_{\mu < \nu} \sigma(u_\mu \, \bar u_\nu)
\prod_{\mu \le \nu} \sigma(u_\mu \, u_\nu) \, Z'_{\mathrm U}(n; \bm u)
\]
which can be also defined as
\[
F_{\mathrm U}(n; \bm u) = (-1)^n \sigma(a)^n \prod_{i, j} \left[
\sigma(u_{2i-1}^3 \, \bar u_{2j}^3) \, \sigma(u_{2i-1}^3 \, u_{2j}^3)
\right] \det M_{\mathrm U}(n; \bm u).
\]
We have the following analogue of Lemmas \ref{l:4}, \ref{l:5} and
\ref{l:6}.

\begin{lemma} \label{l:10}
The function $F_{\mathrm U}(n; \bm u)$ has the following properties.

\begin{itemize}

\item[\rm (a)] For every $\mu = 1, \ldots, 2n$ one has
\begin{eqnarray*}
\lefteqn{F_{\mathrm U}(n; (u_1, \ldots, u_\mu, \ldots, u_{2n}))}
\\[.5em] && {} + F_{\mathrm U}(n; (u_1, \ldots, a^2 u_\mu, \ldots,
u_{2n})) + F_{\mathrm U}(n; (u_1, \ldots, a^4 u_\mu, \ldots, u_{2n})) =
0.
\end{eqnarray*}

\item[\rm (b)] For every $\mu = 1, \ldots, 2n$ the function $u_\mu^{6n-2}
F_{\mathrm U}(n, \bm u)$ is a polynomial of degree $6n-2$ in~$u_\mu^2$.

\item[\rm (c)]
The function $F_{\mathrm U}(n, \bm u)$ turns to zero if for some $\mu
\ne \nu$ either $u^2_\mu = u^2_\nu$, or $u^2_\mu = \bar u^2_\nu$, and if
$u_\mu^4 = 1$ for some $\mu$.

\end{itemize}

\end{lemma}

\begin{lemma} \label{l:11}
The function $F_{\mathrm U}(n, \bm u)$ satisfying Lemma \ref{l:10} is
proportional to the determinant of the matrix
\[
P_{\mathrm U}(n; \bm u) = \left( \begin{array}{ccccc}
\sigma(u_1^{6n-2}) & \sigma(u_2^{6n-2}) & \sigma(u_3^{6n-2}) & \ldots &
\sigma(u_{2n}^{6n-2}) \\[.4em]
\sigma(u_1^{6n-4}) & \sigma(u_2^{6n-4}) & \sigma(u_3^{6n-4}) & \ldots &
\sigma(u_{2n}^{6n-4}) \\[.4em]
\sigma(u_1^8) & \sigma(u_2^{6n-8}) & \sigma(u_3^{6n-8}) & \ldots &
\sigma(u_{2n}^{6n-8}) \\
\vdots & \vdots & \vdots & \ddots & \vdots \\
\sigma(u_1^2) & \sigma(u_2^2) & \sigma(u_3^2) & \dots & \sigma(u_{2n}^2)
\end{array} \right).
\]
\end{lemma}

\begin{proof}
Taking into account the statement (b) of Lemma \ref{l:10}, for some fixed
value of $\mu$ we write
\[
F_{\mathrm U}(n; \bm u) = \sum_{k=1}^{6n - 1} a_k^{(\mu)}(u_1, \ldots,
\hat u_\mu, \ldots, u_{2n}) \, u^{6n - 2k}_\mu.
\]
From the statement (a) of Lemma \ref{l:10} it follows that $a_k^{(\mu)}
= 0$ if $k$ is divisible by 3. Therefore, one has
\[
F_{\mathrm U}(n; \bm u) = \sum_{\scriptstyle k=1 \atop \scriptstyle k \ne 0
\ \mathrm{mod} \ 3}^{6n - 1} a_k^{(\mu)}(u_1, \ldots, \hat u_\mu, \ldots,
u_{2n}) \, u^{6n - 2k}_\mu.
\]
Hence, we have only $4n$ unknown functions $a_k^{(\mu)}$. Using the
statement (c) of Lemma \ref{l:10}, for all $\nu \ne \mu$ one has
\begin{eqnarray*}
&& \sum_{\scriptstyle k=1 \atop \scriptstyle k \ne 0 \ \mathrm{
mod} \ 3}^{6n - 1} a_k^{(\mu)}(u_1, \ldots, \hat
u_\mu, \ldots, u_{2n}) \, u^{6n - 2k}_\nu = 0, \\
&& \sum_{\scriptstyle k=1 \atop \scriptstyle k \ne 0 \ \mathrm{
mod} \ 3}^{6n - 1} a_k^{(\mu)}(u_1, \ldots, \hat u_\mu, \ldots, u_{2n}) \,
\bar u^{6n - 2k}_\nu = 0,
\end{eqnarray*}
and, furthermore,
\[
\sum_{\scriptstyle k=1 \atop \scriptstyle k \ne 0 \ \mathrm{mod} \ 3}^{6n -
1} a_k^{(\mu)}(u_1, \ldots, \hat u_\mu, \ldots, u_{2n}) = 0.
\]
Thus, we have a system of $4n - 1$ linear equations for $4n$ functions
$a_k^{(\mu)}$. In general position the rank of this system is equal to
$4n - 1$. Hence, it determines the functions $a_k^{(\mu)}$ uniquely up
to a common factor which cannot depend on $u_\mu$. Considering all
values of $\mu$, we conclude that Lemma \ref{l:10} determines the
function $F_{\mathrm U}(n, \bm u)$ uniquely up to a constant
factor. It is not difficult to get convinced that the determinant of the
matrix $P_{\mathrm U}(n, \bm u)$ satisfies Lemma \ref{l:10}. Hence, the
function $F_{\mathrm U}(n; \bm u)$ must be proportional to this
determinant.
\end{proof}

\begin{lemma} \label{l:11a}
The determinant of the matrix $P_{\mathrm U}$ satisfies the recurrent
relation
\[
\frac{\det P_{\mathrm U}(n; (u_1, \ldots, u_{2n-2}, u_{2n-1}, a \,
u_{2n-1})}{\det P_{\mathrm U}(n-1; (u_1, \ldots u_{2n-2})} =
\sigma(a) \, \sigma(u_{2n-1}^6) \prod_{\mu=1}^{2n-2} [\sigma(u_\mu \, \bar
u_{2n-1}) \, \sigma(u_\mu \, u_{2n-1})].
\]
\end{lemma}

\begin{proof}
Consider the matrix
\[
P'_{\mathrm U}(n; \bm u) = \left( \begin{array}{cccc}
\sigma(u_1^{6n-2}) & \ldots & \sigma(u_{2n-1}^{6n-2}) &
\sigma(u_{2n}^{6n-2}) \\[.4em]
\sigma(u_1^{6n-8}) & \ldots & \sigma(u_{2n-1}^{6n-8}) &
\sigma(u_{2n}^{6n-8}) \\[.4em]
\sigma(u_1^{6n-14}) & \ldots & \sigma(u_{2n-1}^{6n-14}) &
\sigma(u_{2n}^{6n-14}) \\
\vdots & \ddots & \vdots & \vdots \\
\sigma(u_1^{-6n+10}) & \ldots & \sigma(u_{2n-1}^{-6n+10}) &
\sigma(u_{2n}^{-6n+10}) \\[.4em]
\sigma(u_1^{-6n+4}) & \ldots & \sigma(u_{2n-1}^{-6n+4}) &
\sigma(u_{2n}^{-6n+4}).
\end{array} \right).
\]
One can get convinced that
\[
\det P_{\mathrm U}(n; \bm u) = (-1)^n \det P'_{\mathrm U}(n; \bm u).
\]
For $a = \exp(\rmi \, \pi/3)$ the determinant of the matrix $P'_{\mathrm
U}(n; \bm u)$ is
\[
\left| \begin{array}{cccc}
\sigma(u_1^{6n-2}) & \ldots & \sigma(u_{2n-1}^{6n-2}) & \sigma(\bar a^2 \,
u_{2n-1}^{6n-2}) \\[.4em]
\sigma(u_1^{6n-8}) & \ldots & \sigma(u_{2n-1}^{6n-8}) & \sigma(\bar a^2 \,
u_{2n-1}^{6n-8}) \\
\vdots & \ddots & \vdots & \vdots \\
\sigma(u_1^{-6n+16}) & \ldots & \sigma(u_{2n-1}^{-6n+16}) & \sigma(\bar a^2
\, u_{2n-1}^{-6n+16}) \\[.4em]
\sigma(u_1^{-6n+10}) & \ldots & \sigma(u_{2n-1}^{-6n+10}) & \sigma(\bar a^2
\, u_{2n-1}^{-6n+10}) \\[.4em]
\sigma(u_1^{-6n+4}) & \ldots & \sigma(u_{2n-1}^{-6n+4}) & \sigma(\bar a^2
\, u_{2n-1}^{-6n+4})
\end{array} \right|.
\]
Let us substruct the second row multiplied by $u_{2n-1}^6 + u_{2n-1}^{-6}$
from the first row, and then add the third row. Repeat this procedure
starting from the second row and so on whenever is possible. As the result
we have the determinant
\[
\left| \begin{array}{cccc}
\sigma(u_1^3 \, \bar u_{2n-1}^3) \, \sigma(u_1^3 \, u_{2n-1}^3) \,
\sigma(u_1^{6n-8}) & \ldots & 0 & 0 \\[.4em]
\sigma(u_1^3 \, \bar u_{2n-1}^3) \, \sigma(u_1^3 \, u_{2n-1}^3) \,
\sigma(u_1^{6n-14}) & \ldots & 0 & 0 \\
\vdots & \ddots & \vdots & \vdots \\
\sigma(u_1^3 \, \bar u_{2n-1}^3) \, \sigma(u_1^3 \, u_{2n-1}^3) \,
\sigma(u_1^{-6n+10}) & \ldots & 0 & 0 \\[.4em]
\sigma(u_1^{-6n+10}) & \ldots &
\sigma(u_{2n-1}^{-6n+10}) & \sigma(\bar a^2 \, u_{2n-1}^{-6n+10}) \\[.4em]
\sigma(u_1^{-6n+4}) & \ldots &
\sigma(u_{2n-1}^{-6n+4}) & \sigma(\bar a^2 \, u_{2n-1}^{-6n+4})
\end{array} \right|.
\]
The last expression gives the equality
\[
\frac{\det P'_{\mathrm U}(n; (u_1, \ldots, u_{2n-2}, u_{2n-1}, a \,
u_{2n}))}{\det P'_{\mathrm U}(n-1; (u_1, \ldots, u_{2n-2}))} = -
\sigma(a) \, \sigma(u_{2n-1}^6) \prod_{\mu=1}^{2n-2} [\sigma(u_\mu \, \bar
u_{2n-1}) \, \sigma(u_\mu \, u_{2n-1})]
\]
which implies the statement of the Lemma.
\end{proof}

\begin{theorem} \label{t:3}
For $a = \exp(\rmi \, \pi/3)$ the function $Z'_{\mathrm U}(n; \bm u)$ has
the following determinant representation:
\[
Z'_{\mathrm U}(n; \bm u) = \frac{\sigma(a)^n}{\displaystyle
\prod_{\mu < \nu} \sigma(u_\mu \, \bar u_\nu) \prod_{\mu \le
\nu} \sigma(u_\mu \, u_\nu)} \, \det P_{\mathrm U}(n; \bm u).
\]
\end{theorem}

\noindent Similarly as it was for general ASMs, we can prove the above
theorem using Lemma \ref{l:11a} and the corresponding recurrent relation
for $Z_{\mathrm U}(n; \bm x, \bm y)$ found by Kuperberg \cite{Kup02}.

\begin{corollary}
For $a = \exp(\rmi\, \pi/3)$ the Laurent polynomial
\[
\frac{Z_{\mathrm U}(n; \bm x, \bm y)}{\displaystyle \prod_{i=1}^n
\sigma(y_i \, \bar b) \, \sigma(x_i^2 a^2)}
\]
is symmetric in the union of the coordinates of $\bm x$ and $\bm y$.
\end{corollary}

To treat the refined enumeration problem introduce the function
\[
f_{\mathrm U}(n; u) = \sigma(u)^{4n-2} \, \sigma(u^2)\, Z'_{\mathrm U}(n;
(u, 1, \ldots, 1)).
\]

\begin{lemma} \label{l:12}
The function $f_{\mathrm U}(n; u)$ has the following properties:
\begin{itemize}

\item[\rm (a)] The function $u^{6n-2} f_{\mathrm U}(n; u)$ is a polynomial
of degree $6n-2$ in $u^2$.

\item[\rm (b)] The function $f_{\mathrm U}(n; u)$ satisfies the relation
\[ f(n; \bar u) = - f(n; u).
\]

\item[\rm (c)] The function $f_{\mathrm U}(n; u)$ obeys the equality
\[
f_{\mathrm U}(n; u) + f_{\mathrm U}(n; a^2 u) + f_{\mathrm U}(n; a^4 u) =
0.
\]

\item[\rm (d)] The Laurent polynomial $f_{\mathrm U}(n; u)$ is divisible by
$\sigma(u)^{4n-2}$ and by $\sigma(u^2)$.

\end{itemize}
\end{lemma}
The property (b) follows now, for example, from the determinant
representation for $Z'_{\mathrm U}(n; \bm u)$ supplied by Theorem
\ref{t:3}.
All other properties are evident.

Since $\sigma(u^2) = (u + \bar u) \, \sigma(u)$, the function $f_{\mathrm
U}(n; u)$ is actually divisible by $\sigma(u)^{4n-1}$. Comparing now Lemmas
\ref{l:12} and \ref{l:8} and taking into account Lemma \ref{l:9}, we
conclude that the function $f_{\mathrm U}(n; u)$ is proportional to the
function $\varphi(2n; u)$. Let us find the corresponding coefficient. To
this end consider equality (\ref{9}) for $a = \exp(\rmi\,\pi/\,3)$.
Taking into account an evident property
\begin{equation}
A_{\mathrm U}(2n, 1; 1, y) = y \, A_{\mathrm U}(2n-2; 1, y),
\label{16}
\end{equation}
we obtain the relation
\[
A_{\mathrm U}(2n-2; 1, y) = \frac{Z_{\mathrm U}(n; \bm 1, (a, 1, \ldots,
1))} {\sigma(a)^{2n^2} \, \sigma(a \, b)^{n-1} \sigma(\bar a \, b)},
\]
which after some simple calculations gives
\[
f_{\mathrm U}(n; u) = A_{\mathrm U}(2n-2; 1, y) \frac{\sigma(a)^{2 n^2 + 3n
-1} \, \sigma(a \, b)^{n-1}}{\sigma(b)^{n-1}} \, \varphi(2n; u).
\]
Using this relation, it is not difficult to demonstrate that equality
(\ref{9}) for $a = \exp(\rmi\,\pi/\,3)$ can be written as
\begin{equation}
\frac{1}{A_{\mathrm U}(2n-2; 1, y)} \sum_{r=1}^{2n} A_{\mathrm U}(2n, r; 1,
y) \, t^{r-1} = \frac{\sigma(b \, \bar u) \, \sigma(a)^{6n-2} \,
\varphi(2n; u)}{\sigma(b \, a) \, \sigma(a \, u)^{2n-1} \, \sigma(u)^{4n-2}
\, \sigma(u^2)}. \label{10}
\end{equation}
Comparing (\ref{10}) with (\ref{11}), we come to

\begin{theorem} \label{t:4}
The following equality
\begin{equation}
\frac{1}{A_{\mathrm U}(2n-2; 1, y)} \sum_{r=1}^{2n} A_{\mathrm U}(2n, r; 1,
y) \, t^{r-1} = \frac{1}{A(2n-1)} \, \frac{t+y}{t+1} \sum_{r=1}^{2n} A(2n,
r) \, t^{r-1}, \label{23}
\end{equation}
is valid.
\end{theorem}

The statement of Theorem \ref{t:4} for $t=1$ takes the form
\begin{equation}
\frac{A_{\mathrm U}(2n, 1, y)}{A_{\mathrm U}(2n-2, 1, y)} = \frac{1}{2} \,
(1 + y)\frac{A(2n)}{A(2n-1)}. \label{12}
\end{equation}
For $n = 1$ square ice with U-turn boundary has two states,
\begin{figure}[ht]
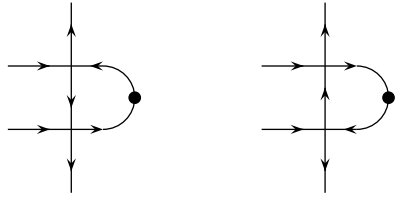

\[
\psset{unit=2em}
\pspicture(1,1)(4,4)
  \arrowLine(1,2)(2,2)
  \arrowLine(1,3)(2,3)
  \psline{->}(2,2)(2.5,2)
  \psline{-<}(2.4,3)(2.5,3)
  \psline(2,3)(2.5,3)
  \arrowLine(2,2)(2,1)
  \arrowLine(2,3)(2,4)
  \arrowLine(2,3)(2,2)
  \psarc(2.5,2.5){.5}{270}{90}
  \qdisk(3,2.5){.1}
\endpspicture
\qquad
\pspicture(1,1)(4,4)
  \arrowLine(1,2)(2,2)
  \arrowLine(1,3)(2,3)
  \psline{-<}(2.4,2)(2.5,2)
  \psline(2,2)(2.5,2)
  \psline{->}(2,3)(2.5,3)
  \arrowLine(2,2)(2,1)
  \arrowLine(2,3)(2,4)
  \arrowLine(2,2)(2,3)
  \psarc(2.5,2.5){.5}{270}{90}
  \qdisk(3,2.5){.1}
\endpspicture
\]
\caption{The states of square ice with U-turn boundary for $n = 1$}
\label{f:uasm1}
\end{figure}
one state with
upward U-turn and one with downward U-turn, see Figure \ref{f:uasm1}.
Therefore, one has
\[
A_{\mathrm U}(2; 1, y) = (1 + y),
\]
and equality (\ref{12}) gives
\[
A_{\mathrm U}(2n; 1, y) = \frac{1}{2^{n}} \, (1 + y)^n \prod_{k=1}^{n}
\frac{A(2k)}{A(2k-1)}.
\]
Using the famous relation
\[
\frac{A(n)}{A(n-1)} = \frac{(3n - 2)! \, (n - 1)!}{(2n - 1)! \, (2n - 2)!},
\]
one can rewrite the above formula for $A_{\mathrm U}(2n; 1, y)$ as
\begin{equation}
A_{\mathrm U}(2n; 1, y) = \frac{1}{2^{n}} \, (1 + y)^n \prod_{k=1}^{n}
\frac{(6k - 2)! \, (2k - 1)!}{(4k - 1)! \, (4k - 2)!}. \label{13}
\end{equation}

\begin{figure}[ht]
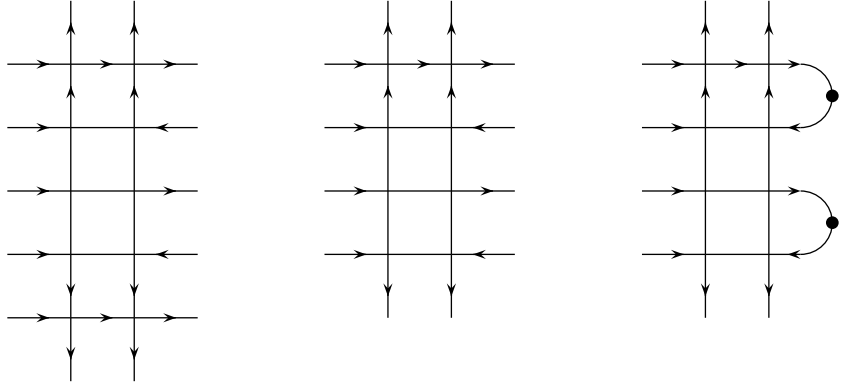

\[
\psset{unit=2em}
\pspicture(0,0)(3,6)
  \arrowLine(0,5)(1,5) \arrowLine(1,5)(2,5) \arrowLine(2,5)(3,5)
  \arrowLine(0,4)(1,4) \psline(1,4)(2,4)    \arrowLine(3,4)(2,4)
  \arrowLine(0,3)(1,3) \psline(1,3)(2,3)    \arrowLine(2,3)(3,3)
  \arrowLine(0,2)(1,2) \psline(1,2)(2,2)    \arrowLine(3,2)(2,2)
  \arrowLine(0,1)(1,1) \arrowLine(1,1)(2,1) \arrowLine(2,1)(3,1)
  \arrowLine(1,5)(1,6) \arrowLine(1,4)(1,5)
  \psline(1,4)(1,2)
  \arrowLine(1,2)(1,1) \arrowLine(1,1)(1,0)
  \arrowLine(2,5)(2,6) \arrowLine(2,4)(2,5)
  \psline(2,4)(2,2)
  \arrowLine(2,2)(2,1) \arrowLine(2,1)(2,0)
\endpspicture
\qquad \qquad
\pspicture(0,0)(3,6)
  \arrowLine(0,5)(1,5) \arrowLine(1,5)(2,5) \arrowLine(2,5)(3,5)
  \arrowLine(0,4)(1,4) \psline(1,4)(2,4)    \arrowLine(3,4)(2,4)
  \arrowLine(0,3)(1,3) \psline(1,3)(2,3)    \arrowLine(2,3)(3,3)
  \arrowLine(0,2)(1,2) \psline(1,2)(2,2)    \arrowLine(3,2)(2,2)
  \arrowLine(1,5)(1,6) \arrowLine(1,4)(1,5)
  \psline(1,4)(1,2)    \arrowLine(1,2)(1,1)
  \arrowLine(2,5)(2,6) \arrowLine(2,4)(2,5)
  \psline(2,4)(2,2)    \arrowLine(2,2)(2,1)
\endpspicture
\qquad \qquad
\pspicture(0,0)(3,6)
  \arrowLine(0,5)(1,5) \arrowLine(1,5)(2,5) \psline{->}(2,5)(2.5,5)
  \arrowLine(0,4)(1,4) \psline(1,4)(2,4)
  \psline(2.5,4)(2,4)  \psline{-<}(2.4,4)(2.5,4)
  \arrowLine(0,3)(1,3) \psline(1,3)(2,3)    \psline{->}(2,3)(2.5,3)
  \arrowLine(0,2)(1,2) \psline(1,2)(2,2)
  \psarc(2.5,2.5){.5}{270}{90}
  \psarc(2.5,4.5){.5}{270}{90}
  \qdisk(3,2.5){.1}
  \qdisk(3,4.5){.1}
  \psline(2.5,2)(2,2)  \psline{-<}(2.4,2)(2.5,2)
  \arrowLine(1,5)(1,6) \arrowLine(1,4)(1,5)
  \psline(1,4)(1,2)    \arrowLine(1,2)(1,1)
  \arrowLine(2,5)(2,6) \arrowLine(2,4)(2,5)
  \psline(2,4)(2,2)    \arrowLine(2,2)(2,1)
\endpspicture
\]
\caption{Transformation of a state of the square ice with VS boundary into
a state of the square ice with U-turn boundary}
\label{f:vsasmuasm}
\end{figure}

Let us discuss now the connection of ASMs with U-turn boundary and
vertically symmetric ASMs. It is not difficult to see that the central
column of a $(2n+1) \times (2n+1)$ VSASM consists of alternating $1$'s and
$-1$'s, and it is the same for any such matrix. Hence, a VSAM is uniquely
characterised by its left-most $(2n+1) \times n$ submatrix. Therefore, one
can associate with a VSASM a state of the square ice $(2n+1) \times n$
region with the boundary condition defined as in at the left side, out at
the top and bottom and alternating boundary at the right side, see the
first picture in Figure~\ref{f:vsasmuasm}. 
Note also that the top and bottom rows of a VSASM are fixed. Therefore, the
orientation of the edges belonging to the corresponding vertices is fixed
as well. This fact is reflected in Figure \ref{f:vsasmuasm}. Actually one
can remove the top and bottom rows from consideration. However, it is
customary to preserve them, and we will obey this agreement.

Any state of square ice with VS boundary can be transformed into a state of
square ice with U-turn boundary via the following two steps. First, we
remove the bottom-most row  of vertices as it is shown on the second
picture in Figure \ref{f:vsasmuasm}. Second, we connect pairwise the
alternating edges on the right side, see the third picture in Figure
\ref{f:vsasmuasm}. Note that the resulting state has only downward oriented
U-turn vertices. It is clear that any state of square ice with U-turn
boundary which has only downward oriented U-turns can be uniquely
transformed into a state of square ice with VS boundary. Thus, the set of
VSASMs can be identified with a subset of VSASMs.

Returning to enumeration problems, write the equality
\begin{equation}
A_{\mathrm V}(2n+1, r) = A_{\mathrm U}(2n, r; 1, 0) \label{22}
\end{equation}
which follows directly from the relation of VSASMs and UASMs discussed just
above. Having this equality in mind, we see that relation (\ref{13})
implies
\begin{equation}
A_{\mathrm V}(2n+1) = \frac{1}{2^{n}} \prod_{k=1}^{n}
\frac{(6k - 2)! \, (2k - 1)!}{(4k - 1)! \, (4k - 2)!}, \label{14}
\end{equation}
and we can write
\begin{equation}
A_{\mathrm U}(2n; 1, y) = (1 + y)^n A_{\mathrm V}(2n + 1).
\label{17}
\end{equation}
This equality is a partial case of the equality
\begin{equation}
A_{\mathrm U}(2n; x, y) = (1 + y)^n A_{\mathrm V}(2n + 1; x) \label{15}
\end{equation}
proved by Kuperberg \cite{Kup02}. Formula (\ref{14}) is equivalent to the
recurrent relation
\[
\frac{A_{\mathrm V}(2n + 1)}{A_{\mathrm V}(2n - 1)} = \frac{\displaystyle
{6n - 2 \choose 2n}}{\displaystyle 2{4n - 1 \choose 2n}}
\]
conjectured by Robbins \cite{Rob}. It can be shown that relation (\ref{14})
is equivalent to the formula for $A_{\mathrm V}(2n+1)$ obtained by
Kuperberg \cite{Kup02}.

Consider now the statement of Theorem \ref{t:4} for general $t$. Rewrite
(\ref{23}) as
\[
A(2n-1) \, (t+1) \sum_{r=1}^{2n} A_{\mathrm U}(2n, r; 1,
y) \, t^{r-1} = A_{\mathrm U}(2n-2; 1, y) \, (t+y) \sum_{r=1}^{2n} A(2n,
r) \, t^{r-1}
\]
and equate coefficients at different powers of $t$. As the result one
obtains the following equalities
\begin{eqnarray}
\lefteqn{A(2n-1) [A_{\mathrm U}(2n, r-1; 1, y) + A_{\mathrm U}(2n, r; 1,
y)]} \nonumber \\[.5em] 
&& \hspace{2em} {} = A_{\mathrm U}(2n-2; 1, y) [A(2n, r-1) + y \, A(2n,
r)], \qquad r = 2, 3, \ldots, 2n. \label{18}
\end{eqnarray}
Using relations (\ref{16}) and (\ref{17}), we obtain the following solution
to this recurrent relation:
\[
A_{\mathrm U}(2n, r; 1, y) = \frac{(1 + y)^{n-1} \, A_{\mathrm V}(2n -
1)}{A(2n - 1)} \left[ y \, A(2n; r) + \sum_{k=1}^{r-1} (-1)^{r+k-1}
(1 - y) A(2n; k) \right].
\]
For $y = 1$ we have
\[
A_{\mathrm U}(2n; r) = 2^{n-1} \frac{A_{\mathrm V}(2n - 1)}{A(2n - 1)} \,
A(2n; r).
\]
Consider, for example, the case of $n = 2$. In this case there are 12 UASMs
depicted in Fi\-gure~\ref{f:uasm2}.
\begin{figure}[ht]
\begin{eqnarray*}
&
\left(
  \psset{unit=2.5em}
  \pspicture[.45](0,0)(1.3,2)
    \rput(.25,.25){$0$}
    \rput(.75,.25){$0$}
    \rput(.25,.75){$0$}
    \rput(.75,.75){$+$}
    \rput(.25,1.25){$0$}
    \rput(.75,1.25){$0$}
    \rput(.25,1.75){$+$}
    \rput(.75,1.75){$0$}
    \psbezier(1,.75)(1.3,.75)(1.3,.25)(1,.25)
	\psbezier(1,1.75)(1.3,1.75)(1.3,1.25)(1,1.25)
  \endpspicture
\right.
\quad
\left(
  \psset{unit=2.5em}
  \pspicture[.45](0,0)(1.3,2)
    \rput(.25,.25){$0$}
    \rput(.75,.25){$+$}
    \rput(.25,.75){$0$}
    \rput(.75,.75){$0$}
    \rput(.25,1.25){$0$}
    \rput(.75,1.25){$0$}
    \rput(.25,1.75){$+$}
    \rput(.75,1.75){$0$}
    \psbezier(1,.75)(1.3,.75)(1.3,.25)(1,.25)
	\psbezier(1,1.75)(1.3,1.75)(1.3,1.25)(1,1.25)
  \endpspicture
\right.
\quad
\left(
  \psset{unit=2.5em}
  \pspicture[.45](0,0)(1.3,2)
    \rput(.25,.25){$0$}
    \rput(.75,.25){$0$}
    \rput(.25,.75){$0$}
    \rput(.75,.75){$+$}
    \rput(.25,1.25){$+$}
    \rput(.75,1.25){$0$}
    \rput(.25,1.75){$0$}
    \rput(.75,1.75){$0$}
    \psbezier(1,.75)(1.3,.75)(1.3,.25)(1,.25)
	\psbezier(1,1.75)(1.3,1.75)(1.3,1.25)(1,1.25)
  \endpspicture
\right.
\quad
\left(
  \psset{unit=2.5em}
  \pspicture[.45](0,0)(1.3,2)
    \rput(.25,.25){$0$}
    \rput(.75,.25){$+$}
    \rput(.25,.75){$0$}
    \rput(.75,.75){$0$}
    \rput(.25,1.25){$+$}
    \rput(.75,1.25){$0$}
    \rput(.25,1.75){$0$}
    \rput(.75,1.75){$0$}
    \psbezier(1,.75)(1.3,.75)(1.3,.25)(1,.25)
	\psbezier(1,1.75)(1.3,1.75)(1.3,1.25)(1,1.25)
  \endpspicture
\right.
\quad
\left(
  \psset{unit=2.5em}
  \pspicture[.45](0,0)(1.3,2)
    \rput(.25,.25){$0$}
    \rput(.75,.25){$0$}
    \rput(.25,.75){$0$}
    \rput(.75,.75){$+$}
    \rput(.25,1.25){$+$}
    \rput(.75,1.25){$-$}
    \rput(.25,1.75){$0$}
    \rput(.75,1.75){$+$}
    \psbezier(1,.75)(1.3,.75)(1.3,.25)(1,.25)
	\psbezier(1,1.75)(1.3,1.75)(1.3,1.25)(1,1.25)
  \endpspicture
\right.
\quad
\left(
  \psset{unit=2.5em}
  \pspicture[.45](0,0)(1.3,2)
    \rput(.25,.25){$0$}
    \rput(.75,.25){$+$}
    \rput(.25,.75){$0$}
    \rput(.75,.75){$0$}
    \rput(.25,1.25){$+$}
    \rput(.75,1.25){$-$}
    \rput(.25,1.75){$0$}
    \rput(.75,1.75){$+$}
    \psbezier(1,.75)(1.3,.75)(1.3,.25)(1,.25)
	\psbezier(1,1.75)(1.3,1.75)(1.3,1.25)(1,1.25)
  \endpspicture
\right. \\[1em]
&
\left(
  \psset{unit=2.5em}
  \pspicture[.45](0,0)(1.3,2)
    \rput(.25,.25){$0$}
    \rput(.75,.25){$0$}
    \rput(.25,.75){$+$}
    \rput(.75,.75){$0$}
    \rput(.25,1.25){$0$}
    \rput(.75,1.25){$0$}
    \rput(.25,1.75){$0$}
    \rput(.75,1.75){$+$}
    \psbezier(1,.75)(1.3,.75)(1.3,.25)(1,.25)
	\psbezier(1,1.75)(1.3,1.75)(1.3,1.25)(1,1.25)
  \endpspicture
\right.
\quad
\left(
  \psset{unit=2.5em}
  \pspicture[.45](0,0)(1.3,2)
    \rput(.25,.25){$0$}
    \rput(.75,.25){$0$}
    \rput(.25,.75){$+$}
    \rput(.75,.75){$0$}
    \rput(.25,1.25){$0$}
    \rput(.75,1.25){$+$}
    \rput(.25,1.75){$0$}
    \rput(.75,1.75){$0$}
    \psbezier(1,.75)(1.3,.75)(1.3,.25)(1,.25)
	\psbezier(1,1.75)(1.3,1.75)(1.3,1.25)(1,1.25)
  \endpspicture
\right.
\quad
\left(
  \psset{unit=2.5em}
  \pspicture[.45](0,0)(1.3,2)
    \rput(.25,.25){$0$}
    \rput(.75,.25){$+$}
    \rput(.25,.75){$+$}
    \rput(.75,.75){$-$}
    \rput(.25,1.25){$0$}
    \rput(.75,1.25){$0$}
    \rput(.25,1.75){$0$}
    \rput(.75,1.75){$+$}
    \psbezier(1,.75)(1.3,.75)(1.3,.25)(1,.25)
	\psbezier(1,1.75)(1.3,1.75)(1.3,1.25)(1,1.25)
  \endpspicture
\right.
\quad
\left(
  \psset{unit=2.5em}
  \pspicture[.45](0,0)(1.3,2)
    \rput(.25,.25){$0$}
    \rput(.75,.25){$+$}
    \rput(.25,.75){$+$}
    \rput(.75,.75){$-$}
    \rput(.25,1.25){$0$}
    \rput(.75,1.25){$+$}
    \rput(.25,1.75){$0$}
    \rput(.75,1.75){$0$}
    \psbezier(1,.75)(1.3,.75)(1.3,.25)(1,.25)
	\psbezier(1,1.75)(1.3,1.75)(1.3,1.25)(1,1.25)
  \endpspicture
\right.
\quad
\left(
  \psset{unit=2.5em}
  \pspicture[.45](0,0)(1.3,2)
    \rput(.25,.25){$+$}
    \rput(.75,.25){$0$}
    \rput(.25,.75){$0$}
    \rput(.75,.75){$0$}
    \rput(.25,1.25){$0$}
    \rput(.75,1.25){$0$}
    \rput(.25,1.75){$0$}
    \rput(.75,1.75){$+$}
    \psbezier(1,.75)(1.3,.75)(1.3,.25)(1,.25)
	\psbezier(1,1.75)(1.3,1.75)(1.3,1.25)(1,1.25)
  \endpspicture
\right.
\quad
\left(
  \psset{unit=2.5em}
  \pspicture[.45](0,0)(1.3,2)
    \rput(.25,.25){$+$}
    \rput(.75,.25){$0$}
    \rput(.25,.75){$0$}
    \rput(.75,.75){$0$}
    \rput(.25,1.25){$0$}
    \rput(.75,1.25){$+$}
    \rput(.25,1.75){$0$}
    \rput(.75,1.75){$0$}
    \psbezier(1,.75)(1.3,.75)(1.3,.25)(1,.25)
	\psbezier(1,1.75)(1.3,1.75)(1.3,1.25)(1,1.25)
  \endpspicture
\right.
\end{eqnarray*}
\caption{All UASMs for $n=2$}
\label{f:uasm2}
\end{figure}
The numbers $A_{\mathrm U}(2n, r)$ are 2, 4, 4, 2. The corresponding
numbers $A(2n, r)$ are 7, 14, 14, 7. One can easy find that $A_{\mathrm
V}(3) = 1$ and $A(3) = 7$. Hence, our relation is valid in this case.

Relation (\ref{18}) for $y = 0$ gives the recurrent relation
\[
A(2n - 1) [A_{\mathrm V}(2n + 1, r - 1) + A_{\mathrm V}(2n + 1, r)] =
A_{\mathrm V}(2n - 1) \, A(2n, r - 1)
\]
which has the solution
\[
A_{\mathrm V}(2n + 1, r) = \frac{A_{\mathrm V}(2n - 1)}{A(2n - 1)}
\sum_{k=1}^{r-1} (-1)^{r+k-1} A(2n, k). 
\]
Using the relation \cite{Zei96a, Str02}
\[
\frac{A(n, r)}{A(n - 1)} = \frac{1}{(2n - 2)!} \frac{(n + r - 2)! \, (2n -
r - 1)!}{(r - 1)! \, (n - r)!},
\]
we come to the equality
\[
A_{\mathrm V}(2n + 1, r) = \frac{A_{\mathrm V}(2n - 1)}{(4n - 2)!}
\sum_{k=1}^{r-1} (-1)^{r+k-1} \frac{(2n + k - 2)! \, (4n - k - 1)!}{(k -
1)! \, (2n - k)!}.
\]
Note that in terms of the generating functions one has the equality
\begin{equation}
\frac{1}{A_{\mathrm V}(2n - 1)} \sum_{r=1}^{2n} A_{\mathrm V}(2n + 1; r) \,
t^{r-1} = \frac{1}{A(2n - 1)} \frac{t}{t + 1} \sum_{r=1}^{2n} A(2n, r) \,
t^{r-1}, \label{21}
\end{equation}
which follows immediately from the statement of Theorem \ref{t:4} if we put
$y = 0$ and take into account equality (\ref{22}).

Let us formulate the main enumeration results of this section as a
corollary of Theorem \ref{t:4}:

\begin{corollary} \label{c}
The following equalities
\begin{eqnarray*}
& \displaystyle A_{\mathrm V}(2n+1) = \frac{1}{2^{n}} \prod_{k=1}^{n}
\frac{(6k - 2)! \, (2k - 1)!}{(4k - 1)! \, (4k - 2)!}, \\[.5em]
& A_{\mathrm U}(2n) = 2^n A_{\mathrm V}(2n + 1), \\[.5em]
& \displaystyle A_{\mathrm U}(2n; r) = 2^{n-1} \frac{A_{\mathrm V}(2n -
1)}{A(2n - 1)} \, A(2n; r), \\[.5em]
& \displaystyle A_{\mathrm V}(2n + 1, r) = \frac{A_{\mathrm V}(2n - 1)}{(4n
- 2)!} \sum_{k=1}^{r-1} (-1)^{r+k-1} \frac{(2n + k - 2)! \, (4n - k -
1)!}{(k - 1)! \, (2n - k)!}
\end{eqnarray*}
are valid.
\end{corollary}

\section{Off-diagonally symmetric ASMs}

\begin{figure}[ht]
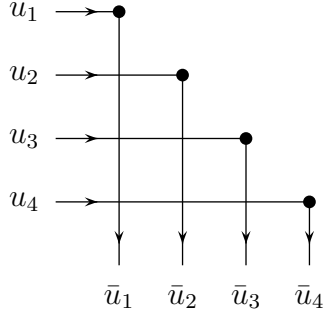

\[
\psset{unit=2em}
\pspicture(0,.3)(5.5,5.5)
  \arrowLine(1,5)(2,5)
  \arrowLine(1,4)(2,4) \psline(2,4)(3,4)
  \arrowLine(1,3)(2,3) \psline(2,3)(4,3)
  \arrowLine(1,2)(2,2) \psline(2,2)(5,2)
  \psline(2,5)(2,2) \arrowLine(2,2)(2,1)
  \psline(3,4)(3,2) \arrowLine(3,2)(3,1)
  \psline(4,3)(4,2) \arrowLine(4,2)(4,1)
  \arrowLine(5,2)(5,1)
  \qdisk(2,5){.1}   \qdisk(3,4){.1}  \qdisk(4,3){.1}  \qdisk(5,2){.1}
  \rput(.5,5){$u_1$}
  \rput(.5,4){$u_2$}
  \rput(.5,3){$u_3$}
  \rput(.5,2){$u_4$}
  \rput(2,.5){$\bar u_1$}
  \rput(3,.5){$\bar u_2$}
  \rput(4,.5){$\bar u_3$}
  \rput(5,.5){$\bar u_4$}
\endpspicture
\]
\caption{Square ice with OS boundary}
\label{f:osasm}
\end{figure}

An off-diagonally symmetric ASM (OSASM) is an ASM which coincides with its
transpose and has null diagonal. An example of the corresponding square ice
model pattern is given in Figure \ref{f:osasm}.
This model was posed by Kuperberg \cite{Kup02} who also found the following
Pfaffian representation for the partition function:
\[
Z_{\mathrm O}(n; \bm u) = \frac{\sigma(a^2)^n \displaystyle \prod_{\mu <
\nu} \alpha(u_\mu \, u_\nu)}{\displaystyle \prod_{\mu < \nu} \sigma(\bar
u_\mu \, u_\nu)} \, \Pf M_{\mathrm O}(n;
\bm u),
\]
where $M_{\mathrm O}(n, \bm x)$ is $2n \times 2n$ matrix with the matrix
elements given by
\[
M_{\mathrm O}(n; \bm u)_{\mu \nu} = \frac{\sigma(\bar u_\mu \,
u_\nu)}{\alpha(u_\mu \, u_\nu)}.
\]
It is not difficult to obtain the relation
\begin{equation}
\sum_{r=2}^{2n} A_{\mathrm O}(2n, r; x) \, t^{r-2} = \frac{Z_{\mathrm O}(n;
(\bar u, 1, \ldots, 1))}{\sigma(a)^{2n^2 - 2n + 1} \, \sigma(a^2)^n \,
\sigma(a \, u)^{2n - 2}}, \label{20}
\end{equation}
where
\[
x = \left[ \displaystyle \frac{\sigma(a^2)}{\sigma(a)} \right]^2, \qquad t
= \frac{\sigma(a \, \bar u)}{\sigma(a \, u)}.
\]

Assume again that $a = \exp(\rmi \pi/3)$. In this case, using equality
(\ref{8}), we write the partition function as
\[
Z_{\mathrm O}(n; \bm u) = (-1)^n \frac{\sigma(a^2)^n \displaystyle
\prod_{\mu < \nu} \sigma(u_\mu^3 \, u_\nu^3) \prod_\mu
\sigma(u_\mu^2)}{\displaystyle \prod_{\mu < \nu} \sigma(\bar
u_\mu \, u_\nu) \prod_{\mu \le \nu} \sigma(u_\mu \, u_\nu)} \,
\Pf M_{\mathrm O}(n; \bm u).
\]
Introduce the function
\[
F_{\mathrm O}(n; \bm u) = \sigma(a)^{-n} \prod_{\mu < \nu} \sigma(\bar
u_\mu \, u_\nu) \prod_{\mu \le \nu} \sigma(u_\mu \, u_\nu) \, Z_{\mathrm
O}(n; \bm u),
\]
which can be equivalently defined by
\[
F_{\mathrm O}(n; \bm u) = (-1)^n \prod_{\mu < \nu} \sigma(u_\mu^3 \,
u_\nu^3) \prod_\mu \sigma(u_\mu^2) \, \Pf M_{\mathrm O}(n; \bm u).
\]

\begin{lemma} \label{l:13}
The function $F_{\mathrm O}(n; \bm u)$ has the following properties.

\begin{itemize}

\item[\rm (a)] For every $\mu = 1, \ldots, 2n$ one has
\begin{eqnarray*}
\lefteqn{F_{\mathrm O}(n; (u_1, \ldots, u_\mu, \ldots, u_{2n}))}
\\[.5em] && {} + F_{\mathrm O}(n; (u_1, \ldots, a^2 u_\mu, \ldots,
u_{2n})) + F_{\mathrm O}(n; (u_1, \ldots, a^4 u_\mu, \ldots, u_{2n})) =
0.
\end{eqnarray*}

\item[\rm (b)] For every $\mu = 1, \ldots, 2n$ the function $u_\mu^{6n-2}
F_{\mathrm O}(n, \bm u)$ is a polynomial of degree $6n-2$ in~$u_\mu^2$.

\item[\rm (c)]
The function $F_{\mathrm O}(n, \bm u)$ turns to zero if for some $\mu
\ne \nu$ either $u^2_\mu = u^2_\nu$, or $u^2_\mu = \bar u^2_\nu$, and if
$u_\mu^4 = 1$ for some $\mu$.

\end{itemize}

\end{lemma}

\begin{proof}
To prove the statement (a) note that for $a = \exp(\rmi \, \pi / 3)$ one
has
\[
M_{\mathrm O}(n; \bm u)_{\mu \nu} = - \frac{\sigma(\bar u_\mu \, u_\nu) \,
\sigma(u_\mu \, u_\nu)}{\sigma(u_\mu^3 \, u_\nu^3)}.
\]
One can get convinced that
\[
\sigma(u_\mu^2) \, M_{\mathrm O}(n; \bm u)_{\mu \nu} = -
\frac{\sigma(u_\mu^2) \, \sigma(u_\nu^2) - \sigma(u_\mu^4)}{\sigma(u_\mu^3
\, u_\nu^3)}.
\]
Taking into account the equality
\[
\sigma(x) + \sigma(a^8 x) + \sigma (a^{16} x) = 0,
\]
which follows from (\ref{19}), one obtains
\begin{eqnarray*}
\lefteqn{\sigma(u_1^2) \, M_{\mathrm O}(n; (u_1, \ldots, u_{2n})_{1 \nu}}
\\[.5em]
&& {} + \sigma(a^4 u_1^2) \, M_{\mathrm O}(n; (a^2 u_1, \ldots, u_{2n}))_{1
\nu} + \sigma(a^8 u_1^2) \, M_{\mathrm O}(n; (a^4 u_1, \ldots, u_{2n}))_{1
\nu} = 0.
\end{eqnarray*}
Since $\Pf M_{\mathrm O}(n; \bm u)$ for every $\nu$ linearly depends on
$M_{\mathrm O}(n; \bm u)_{1 \nu}$, the above equality implies that
\begin{eqnarray*}
\lefteqn{\sigma(u_1^2) \, \Pf M_{\mathrm O}(n; (u_1, \ldots, u_{2n})}
\\[.5em]
&& {} + \sigma(a^4 u_1^2) \Pf \, M_{\mathrm O}(n; (a^2 u_1, \ldots,
u_{2n})) +
\sigma(a^8 u_1^2) \, \Pf M_{\mathrm O}(n; (a^4 u_1, \ldots, u_{2n})) = 0.
\end{eqnarray*}
We see that the statement (a) of the lemma is valid for $\mu=1$. For all
other values of $\mu$ the proof is the same. The statements (b) and (c) are
evident.
\end{proof}

\noindent Comparing Lemmas \ref{l:13} and \ref{l:10}, and taking into
account Lemma \ref{l:11}, we see that the function $F_{\mathrm O}(n; \bm
u)$ is proportional to $\det P_{\mathrm U}(n; \bm u)$.

\begin{theorem} \label{t:5}
For $a = \exp(\rmi \, \pi/3)$ the function $Z_{\mathrm O}(n; \bm u)$ has
the following determinant representation:
\[
Z_{\mathrm O}(n; \bm u) = \frac{\sigma(a)^n}{\displaystyle
\prod_{\mu < \nu} \sigma(u_\mu \, \bar u_\nu) \prod_{\mu \le
\nu} \sigma(u_\mu \, u_\nu)} \, \det P_{\mathrm U}(n; \bm u).
\]
\end{theorem}

\noindent The proof of Theorem \ref{t:5} is based on usage of Lemma
\ref{l:11a} and the recurrent relation for the partition function
$Z_{\mathrm O}(n; \bm u)$ found by Kuperberg \cite{Kup02}.

\begin{corollary}
The partition function $Z_{\mathrm O}(n; \bm u)$ coincides with the
modified partition function $Z'_{\mathrm U}(n; \bm u)$.
\end{corollary}

To treat enumeration problems define the function
\[
f_{\mathrm O}(n; u) = \sigma(u)^{4n - 2} \sigma(u^2) Z_{\mathrm O}(n; (u,
1, \ldots, 1)).
\]
The function $f_{\mathrm O}(n; u)$ has the properties of the function
$f_{\mathrm U}(n; u)$ described by Lemma \ref{l:12}. Therefore, it is
proportional to the function $\varphi(2n; u)$. As for the case of UASMs we
find
\[
f_{\mathrm O}(n; u) = A_{\mathrm O}(2n - 2) \sigma(a)^{2n^2 - 5n - 2}
\varphi(2n; u).
\]
From this equality and relation (\ref{20}) it follows that for $a =
\exp(\rmi \pi/3)$ one has
\[
\frac{1}{A_{\mathrm O}(2n - 2)} \sum_{r=2}^{2n} A_{\mathrm O}(2n, r) \,
t^{r-2} = \frac{\sigma(a)^{6n - 3} \, \varphi(2n; u)}{\sigma(a \, u)^{2n-2}
\, \sigma(u)^{4n-2} \, \sigma(u^2)}.
\]
Comparing this equality with (\ref{11}), we come to

\begin{theorem}
The refined enumerations of OSASMs and ASMs are connected by the equality
\[
\frac{1}{A_{\mathrm O}(2n - 2)} \sum_{r=1}^{2n} A_{\mathrm O}(2n; r) \,
t^{r-1} = \frac{1}{A(2n - 1)} \frac{t}{t + 1} \sum_{r=1}^{2n} A(2n, r) \,
t^{r-1}.
\]
\end{theorem}

Comparing the statement of this theorem with equality (\ref{21}), and
having in mind that
\[
A_{\mathrm O}(2) = A_{\mathrm V}(3) = 1,
\]
we obtain
\begin{corollary}
The refined enumerations of OSASMs and VSASMs coincide:
\[
A_{\mathrm O}(2n, r) = A_{\mathrm V}(2n + 1, r).
\]
\end{corollary}
This is the conjecture made by Kutin and Yuen.

Note that the determinant representaions of Theorems \ref{t:2}, \ref{t:3}
and \ref{t:5} can be obtained from the determinant representaions by
Izergin--Korepin and Kuperberg using some of the equalities between
determinants and Pffafians found by Okada \cite{Oka98}.

{\it Acknowledgments}
The work was supported in part by the Russian
Foundation for Basic Research under grant \# 01--01--00201 and by the INTAS
under grant
\# 00--00561.


\begin{thebibliography}{XX}

\bibitem{BGN01}
M.~T.~Batchelor, J.~de~Gier and B. Nienhuis,
{\em The quantum symmetric XXZ chain at $\Delta=-1/2$, alternating sign
matrices and plane partitions\/},
J. Phys. A: Math. Gen. {\bf 34} (2001) L265--L270;\\
{\tt arXiv:cond-mat/0101385}.

\bibitem{BGN02}
M.~T.~Batchelor, J.~de~Gier, B.~Nienhuis,
{\em The rotor model and combinatorics\/},
Int. J. Mod. Phys. {\bf B16} (2002) 1883--1890;\\
{\tt arXiv:math-ph/0204002}.

\bibitem{EKLP92}
N.~Elkies, G.~Kuperberg, M.~Larsen, and J.~Propp,
{\em Alternating sign matrices and domino tilings, I\/},
J. Algebraic Combin. {\bf 1} (1992) 111--132.

\bibitem{FZZ03}
P.~Di~Francesco, P.~Zinn--Justin and J.--B. Zuber,
{\em A bijection between classes of fully packed loops and plane
partitions\/},\\
{\tt arXiv:math.CO/0311220}.

\bibitem{GBNM01}
J.~de~Gier, M.~Batchelor, B.~Nienhuis and S.~Mitra,
{\em The XXZ spin chain at $\Delta= - 1/2$: Bethe roots, symmetric
functions and determinants\/},\\
{\tt arXiv:math-ph/0110011}.

\bibitem{GNPR03}
J.~de~Gier, B.~Nienhuis, P.~A.~Pearce and V. Rittenberg,
{\em The raise and peel model of a fluctuating interface\/},\\
{\tt arXiv:cond-mat/0301430}.

\bibitem{Ize87}
A.~G.~Izergin,
{\em Partition function of the six-vertex model in a finite volume\/},
Sov. Phys. Dokl. {\bf 32} (1987) 878--879.

\bibitem{Kor82}
V.~E.~Korepin,
{\em Calculation of norms of Bethe wave functions\/},
Commun. Math. Phys. {\bf 86} (1982) 391--418.

\bibitem{KIB93}
V.~E.~Korepin, N.~M.~Bogoliubov, and A.~G.~Izergin,
{\em Quantum Inverse Scattering Method, Correlation Functions
and  Algebraic Bethe Ansatz\/}, 
Cambridge Univ. Press, New York, 1993.

\bibitem{Kup96}
G.~Kuperberg,
{\em Another proof of the alternating-sign matrix conjecture\/},
Int. Math. Res. Notes {\bf 3} (1996) 139--150.

\bibitem{Kup02}
G.~Kuperberg,
{\em Symmetry classes of alternating-sign matrices under one roof\/},
Ann. Math. {\bf 156} (2002) 835--866;\\
{\tt arXiv:math.CO/0008184}.

\bibitem{Ku}
G.~Kuperberg, Domino Forum, 08/29/2003.

\bibitem{MRR82}
W.~H.~Mills, D.~P.~Robbins, and H.~Rumsey,
{\em Proof of the Macdonald conjecture\/},
Invent. Math. {\bf 66} (1982) 73--87.

\bibitem{MRR83}
W.~H.~Mills, D.~P.~Robbins, and H.~Rumsey,
{\em Alternating-sign matrices and descending plane partitions\/},
J. Combin. Theory Ser. A {\bf 34} (1983) 340--359.

\bibitem{Oka98}
S.~Okada,
{\em Application of minor summation formula to rectangular-shaped
representations of classical groups\/},
J. Algebra, {\bf 205} (1998) 337--367.

\bibitem{PRG01}
P.~A.~Pearce, V.~Rittenberg and J.~de~Gier,
{\em Critical Q=1 Potts model and Temperley--Lieb stochastic
processes\/},\\
{\tt arXiv:cond-mat/0108051}.

\bibitem{RSt01a}
A.~V.~Razumov and Yu.~G.~Stroganov,
{\em Spin chains and combinato\-rics\/},
J. Phys. A: Math. Gen. {\bf 34} (2001) 3185--3190;\\
{\tt arXiv:cond-mat/0012141}.

\bibitem{RSt01b}
A.~V.~Razumov and Yu.~G.~Stroganov,
{\em Spin chains and combinatorics: twis\-ted boundary conditions\/},
J. Phys. A: Math. Gen. {\bf 34} (2001) 5335--5340;\\
{\tt arXiv:cond-mat/0102247}.

\bibitem{RSt01c}
A.~V.~Razumov and Yu.~G.~Stroganov,
{\em Combinatorial nature of ground state vector of {\rm O(1)} loop
model\/},\\
{\tt arXiv:math.CO/0104216}.

\bibitem{RSt01d}
A.~V.~Razumov and Yu.~G.~Stroganov,
{\em {\rm O(1)} loop model with different boundary conditions and symmetry
classes of alternating-sign matrices\/},\\
{\tt arXiv:cond-mat/0108103}.

\bibitem{Rob}
D.~P.~Robbins,
{\em Symmetry classes of alternating sign matrices\/},\\
{\tt arXiv:math.CO/0008045}.

\bibitem{Str01}
Yu.~G.~Stroganov,
{\em The importance of being odd\/},
J. Phys. A {\bf 34} (2001) L179-L185

\bibitem{Str02}
Yu.~G.~Stroganov,
{\em A new way to deal with Izergin--Korepin determinant at root of
unity\/},\\
{\tt arXiv:math-ph/0204042}.

\bibitem{Tsu}
O.~Tsuchiya,
{\em Determinant formula for the six-vertex model with reflecting end\/},
J. Math. Phys. {\bf 39} (1998) 5946--5951.

\bibitem{Zei96}
D.~Zeilberger,
{\em Proof of the  alternating sign matrix conjecture\/},
Elec. J. Comb. {\bf 3(2)} (1996) \#R13;\\
{\tt arXiv:math.CO/9407211}.

\bibitem{Zei96a}
D.~Zeilberger,
{\em Proof of the refined alternating sign matrix conjecture\/},
New York J. Math. {\bf 2} (1996) 59--68;\\
{\tt arXiv:math.CO/9506224}.

\bibitem{Zub03}
J.--B. Zuber,
{\em On the counting of fully packed loop configurations. Some new
conjectures\/},\\
{\tt arXiv:math-ph/0309057}.

\end{thebibliography}
\end{document}